\documentclass[aps,pra,onecolumn,superscriptaddress]{revtex4-2}
\usepackage{xcolor}

\usepackage{amsmath,amssymb,graphicx,bm}
\usepackage{microtype}
\usepackage{booktabs}
\usepackage[colorlinks=true, allcolors=blue]{hyperref}
\usepackage{orcidlink}
\usepackage[T1]{fontenc}

\newcommand{\Tr}{\operatorname{Tr}}
\newcommand{\ket}[1]{\lvert#1\rangle}
\newcommand{\bra}[1]{\langle #1\rvert}
\newcommand{\proj}[1]{\lvert#1\rangle\langle #1\rvert}
\newcommand{\E}{{\cal E}}

\newcommand{\dn}{d\vec n}
\newcommand{\dvec}[1]{\lvert #1\rangle\!\rangle}
\newcommand{\dbra}[1]{\langle\!\langle #1\rvert}

\begin{document}
\title{Full-state information-disturbance tradeoff for direction
  estimation with antiparallel spin-coherent pairs}
\author{Massimiliano F. Sacchi\,\orcidlink{0000-0002-8909-2196}}
	\affiliation{CNR-Istituto di Fotonica e Nanotecnologie, Piazza Leonardo da Vinci 32, I-20133, Milano, Italy}
	\affiliation{Dipartimento di Fisica, Università degli Studi di
          Pavia, Via Agostino Bassi 6, I-27100, Pavia, Italy}
\begin{abstract}
We determine the optimal information-disturbance tradeoff for
estimating an unknown spatial direction encoded in two antiparallel
spins. Rotational covariance reduces the optimization over all
instruments to a finite-dimensional Choi problem: a positive seed
operator obeys one trace constraint for each irreducible sector of the
input representation, while both the directional score and the
operation fidelity are linear functionals of this seed. For two
antiparallel spin-$1/2$ particles, whose physical representation
decomposes as $0\oplus1$, we derive the two-multiplier dual problem
and characterize the optimal instrument from the kernel vectors of the
dual slack operator. The optimal operation is a covariant filter with
scalar-vector coherence and is generally not a convex interpolation
between the identity channel and a measure-and-reprepare strategy. At
maximum information we recover the Gisin-Popescu score, but the least
disturbing output state is optimized independently, giving a smaller
disturbance than both the parallel-spin benchmark and antiparallel
measure-and-reprepare. We also formulate the parallel benchmark and, as a
central extension of the method, treat antiparallel spin-coherent states of
arbitrary spin $j$. In this case  the signal coherently occupies all sectors
$\ell=0,\ldots,2j$ of $j\otimes j$, the endpoint information is governed by nearest-neighbor
sector coherences, and the endpoint disturbance is obtained from an
explicit finite block-diagonal eigenvalue problem.
\end{abstract}
\maketitle
\section{Introduction}
A quantum measurement has two inseparable roles.  It produces
classical information about the system, but it also implements a
physical operation on the postmeasurement state.  In general these two
roles are in tension: a more informative measurement causes a larger
back action on the state left for subsequent use.  This tension,
already present in the early discussion of measurement disturbance
\cite{Heisenberg}, is commonly quantified through
information-disturbance tradeoffs.  Such tradeoffs have been studied
for state discrimination and estimation, weak and partial
measurements, continuous variables, and reversible or approximately
reversible measurement schemes
\cite{Helstrom,FuchsPeres,BanaszekPRL,BanaszekDevetak,FuchsJacobs,Ozawa,MistaFilip,Andersen,SacchiMaxEnt,BuscemiSacchi,SacchiSCS,LeeInfoBalance,HongInfoTradeoff,SaberianSharpness,AsadianCovariant}.
Here we consider a geometric version of the problem: the unknown
parameter is a spatial direction encoded in a pair of spins.

Direction estimation is a natural setting for covariant quantum
estimation.  Alice chooses a unit vector uniformly on the sphere and
prepares a state in the corresponding group orbit.  Bob applies an
instrument whose outcome gives a guess for the direction and whose
conditional output is the state that remains after the measurement.
We quantify the information by the standard directional score $s(\vec
n,\vec m)=(1+\vec n\cdot\vec m)/2$, where $\vec m$ is the true
direction and $\vec n$ is the guess
\cite{Holevo,MassarPopescu,Derka,Latorre,GisinPopescu,PeresScudo,Bagan}.
We quantify the disturbance by $D=1-F$, where $F$ is the full-state
operation fidelity between the input pure state of the complete two-spin
signal and the conditional output.  Thus the identity channel has
$D=0$, while an informative measurement generally has $D>0$.  Because
$F$ depends on the output state and not only on the measurement
probabilities, optimizing the disturbance requires optimizing the full
instrument.

A related information-disturbance problem for antiparallel spin-$1/2$
direction transmission was considered in
Ref.~\cite{ZhangAntiTradeoff}.  There, however, the disturbance
functional was marginal: it tested the survival of one output spin
through an observable of the form $|\psi _g\rangle\langle \psi _g
|\otimes I$.  In the present work we use the full-state operation
fidelity of the complete two-spin signal.  This changes the
operational problem: the zero-disturbance point is the identity
operation with random directional score $I=1/2$, rather than the
marginal zero-disturbance point $I=2/3$. The two analyses are
therefore operationally distinct.

The physical comparison motivating the paper is between parallel and
antiparallel encodings.  In both cases the physical rotation acting on
the two spins is the same, namely $V_g=U_g\otimes U_g$, where $U_g$
denotes an irreducible representation of the rotation $g$. The
distinction between the two encodings is the choice of fiducial state.
For two spin-$1/2$ the parallel pair is generated from
$\ket{\uparrow}\ket{\uparrow}$ and is confined to the symmetric
spin-one sector.  The antiparallel pair is generated from
$\ket{\uparrow}\ket{\downarrow}$ and has coherent components in both
the singlet and triplet sectors, i.e. in the decomposition
$1/2\otimes1/2\simeq0\oplus1$.  This sector coherence is the relevant
resource for direction estimation.

The role of collective measurements in extracting information from
multipartite quantum systems was already highlighted by Peres and
Wootters \cite{PeresWootters}.  In the present direction-estimation
setting, Gisin and Popescu showed that two antiparallel spin-$1/2$
particles can carry more directional information than two parallel
ones when collective measurements are allowed \cite{GisinPopescu}.
The advantage is not due to entanglement of the signal state, which is
a product state, but to the coherent occupation of inequivalent
irreducible sectors.  Related developments include optimal
reference-frame transmission, covariant maximum-likelihood
measurements, group-transformation estimation, photonic orienteering
experiments, and recent work on incompatibility and device estimation
with parallel and antiparallel spins \cite{Bartlett,chiri2,chir1,chiri3,ChiribellaRefFrame,TangOrienteering,PatraIncompatibility}.

The question addressed here is whether the antiparallel advantage
survives when measurement back action is included.  At the fully
informative endpoint for
two spin-$1/2$ particles we find that it does: the antiparallel encoding gives a
larger information score and a smaller minimum disturbance than the
parallel encoding.  Moreover, the least disturbing endpoint operation
is not the naive measure-and-reprepare channel that prepares the
guessed antiparallel state.  At fixed optimal measurement, the output
state remains a variational degree of freedom; optimizing it reduces
the disturbance below both the parallel endpoint value and the
antiparallel measure-and-reprepare value.  This separation between
optimal statistics and optimal postmeasurement state is analogous to
what occurs in state-discrimination tradeoffs
\cite{BuscemiSacchi}. The advantage of antiparallel encoding survives
along the entire optimal information-disturbance tradeoff curve.

The main technical tool used in this paper is covariance.  Since the
prior, the signal ensemble, and the score are rotationally invariant,
any strategy can be group averaged without changing the averaged
information or operation fidelity.  The optimization can therefore be
restricted to covariant instruments.  In the Jamio\l{}kowski
representation a covariant instrument is described by one positive
seed Choi operator.  Trace preservation gives one linear constraint
for each irreducible sector of the input representation, and both
figures of merit are linear functionals of the seed.  The tradeoff
boundary is therefore a semidefinite optimization problem with a
simple representation-theoretic structure \cite{bv,WatrousTQI,HayashiGroup}.  We use the standard
formalism of quantum operations, instruments, covariant measurements,
and Choi-Jamio\l{}kowski operators
\cite{chiri3,DaviesLewis,DaviesOpen,OzawaCont,HolevoImprimitivity,HolevoCovInstr,Kraus,Jamio,Choi,CarmeliCovInstr}. The
same framework is closely related to earlier uses of covariance,
vectorization, and group representation theory in Bell measurements,
optimal realizations of positive maps, maximum-likelihood estimation,
and information-disturbance tradeoffs
\cite{lop,bu,SacchiMaxEnt,SacchiSCS,BuscemiSacchi}.  For two
antiparallel spin-$1/2$ particles the supporting-line optimization has
two trace multipliers.  We derive the corresponding dual slack
operator and characterize the optimal seed by its kernel.  For generic
interior points this kernel is one-dimensional, giving a rank-one
covariant filter with scalar-vector mixing.  This mixing is precisely
the structure that exploits the coherence between inequivalent
angular-momentum sectors and explains why the optimal tradeoff is not
a convex interpolation between the identity channel and a
measure-and-reprepare strategy.  We also derive endpoint formulas for
antiparallel spin-coherent states of arbitrary spin $j$, in a setting
closely related to optimal spin-direction encoding and decoding
\cite{PeresScudo,BaganSpinDirection}.  In this case the representation
decomposes as $j\otimes j\simeq\bigoplus_{\ell=0}^{2j}\ell$, and the
fiducial antiparallel state has coherent amplitudes in all sectors.
The maximum information is controlled by nearest-neighbor coherences
between sectors $\ell$ and $\ell+1$, while the minimum endpoint
disturbance is the largest eigenvalue of an explicit finite
block-diagonal operator.  This sector structure is related to the
broader role of finite reference frames, covariance, and approximate
symmetry constraints in quantum information processing and error
correction
\cite{PopescuQRF,PopescuSeparateFrames,HaydenRefFrameQEC,ZhouCovariantQEC,YangBoundedFrames,KongCovariantCodes,MarvianSymmetryLocality,LiuApproxSymmetry,MarvianRotInvCircuits}.  This
arbitrary-spin endpoint analysis is more than a formal generalization:
it shows that the same covariance-and-kernel-vector method organizes
the tradeoff for an increasing number of irreducible sectors,
replacing the two-sector spin-$1/2$ problem by a controlled finite
family of sector constraints and block eigenvalue problems.

The paper is organized as follows.  Section~\ref{sec:covariant}
introduces quantum instruments, the information and disturbance
figures of merit, and the covariant Jamio\l{}kowski reduction.
Sections~\ref{sec:anti_half} and \ref{sec:optimal_method} solve the
antiparallel spin-$1/2$ problem from the covariant dual, while
Sec.~\ref{sec:endpoints} discusses the endpoints and the improvement
over measure-and-reprepare.  Section~\ref{sec:parallel_benchmark}
gives the irreducible parallel-spin benchmark.  The extension to
arbitrary antiparallel spin is formulated in
Sec.~\ref{sec:higher_spin}, and the maximum-information endpoint for
arbitrary spin is evaluated in
Sec.~\ref{sec:endpoints_arbitrary_j}. Appendix~\ref{app:zhang_relation}
briefly relates the present full-state disturbance to the
marginal-disturbance tradeoff of Ref.~\cite{ZhangAntiTradeoff}.
Appendix~\ref{app:fixed_povm_Q} contains the fixed-POVM output
optimization used at the endpoints. Appendices
\ref{app:rank_one_simple_kernel} and~\ref{app:rank_bound} collect the
rank statements for optimal Choi seeds: the one-dimensional-kernel
criterion for rank-one optimality and the general convex-geometric
low-rank bound. Appendix~\ref{asym} derives the large-spin asymptotics
of the antiparallel endpoint problem.

\section{Covariant formulation}
\label{sec:covariant}
\subsection{Instruments and figures of merit}
A measurement with outcome $r$ is described by a quantum instrument,
namely a collection of trace-decreasing completely positive maps \cite{DaviesLewis,OzawaCont,Kraus},
\begin{equation}
        \E_r(\rho)=\sum_\mu A_{r\mu}\rho A_{r\mu}^{\dagger} .
        \label{eq:instrument}
\end{equation}
For a discrete outcome set, the probability of outcome $r$ on an input
state $\rho$ is $p_r=\Tr[\E_r(\rho)]=\Tr[\Pi_r \rho]$, and, when $p_r>0$, the
conditional output state is $\rho_r=\E_r(\rho)/p_r$.  The associated
POVM is
\begin{equation}
        \Pi_r=\sum_\mu A_{r\mu}^{\dagger}A_{r\mu},
        \qquad \sum_r\Pi_r=I_{\rm in} ,
        \label{eq:povm_general}
\end{equation}
where $I_{\rm in}$ denotes the identity operator in the input Hilbert
space.  The completeness relation is equivalent to trace preservation of the
nonselective operation $\sum_r\E_r$.

In the direction-estimation problem the outcomes are continuous and are
identified with unit vectors $\vec n$ on the sphere.  We use the
normalized invariant measure $\int\dn=1$, so that
\begin{equation}
        \int\dn\sum_\mu A_{\vec n\mu}^{\dagger}A_{\vec n\mu}=I _{\rm in} .
        \label{eq:continuous_completeness}
\end{equation}
All probabilities below are therefore probability densities with respect
to $\dn$.

Let the signal states form a covariant orbit
\begin{equation}
        \ket{\psi_g}=V_g\ket{\psi_0},
        \qquad g\in G,
        \label{eq:orbit_general}
\end{equation}
where $G$ is compact and the prior is the normalized Haar measure.  In
the applications below $G={\rm SU}(2)$, or equivalently ${\rm SO}(3)$ for
the orbit of directions, and the fiducial state is associated with the
reference direction $\vec z$.  Although we often write group elements
$g$, the physical outcome is the direction $\vec n=g\vec z$, namely the
coset $gH\in {\rm SU}(2)/{\rm U}(1)$; all outcome integrals are
normalized sphere integrals unless explicitly stated otherwise.
Throughout the paper, $V_g$ denotes the physical representation acting on the
spin system under consideration.  For a pair of physical spin-$j$ particles this
means $V_g=U_g^{(j)}\otimes U_g^{(j)}$, where $U_g^{(j)}$ is the spin-$j$
irreducible representation of the rotation $g$.

For a true reference direction $\vec z$, the information extracted by the
instrument is defined as the average score
\begin{equation}
        I=\int\dn\sum_\mu s(n_z)
        \bra{\psi_z}A_{\vec n\mu}^{\dagger}A_{\vec n\mu}\ket{\psi_z},
        \label{eq:Idef}
\end{equation}
where $s(n_z)=(1+n_z)/2$. This is the standard directional score used
in covariant estimation of a spatial direction
\cite{Holevo,MassarPopescu,GisinPopescu}. For a true direction $\vec
m$ and a guessed direction $\vec n$, more generally $s(\vec n,\vec
m)=(1+\vec n\cdot\vec m)/2 =\cos^2(\theta /2)$, where $\theta$ is the
angle between the two directions.  The score is normalized between
zero and one, gives unit score for a correct guess, zero score for the
opposite direction, and Haar average $1/2$ for a random guess.  For
spin-one-half coherent states it coincides with the transition
probability between the two pure states whose Bloch vectors are $\vec
n$ and $\vec m$.  We use the same score for all spin values because
the estimated parameter is the spatial direction itself.

The disturbance is one minus the operation fidelity,
\begin{equation}
        D=1-F,
        \qquad
        F=\int\dn\sum_\mu
        \left|\bra{\psi_z}A_{\vec n\mu}\ket{\psi_z}\right|^2 .
        \label{eq:Ddef}
\end{equation}
For later use it is helpful to spell out how Eq.~(\ref{eq:Ddef}) is obtained
from the conditional states.  If the input is the pure state
$\rho_z=\proj{\psi_z}$, the probability density of the outcome $\vec n$ is
\begin{equation}
        p(\vec n|z)=\sum_\mu
        \bra{\psi_z}A_{\vec n\mu}^{\dagger}A_{\vec n\mu}\ket{\psi_z} .
\end{equation}
The corresponding conditional output is
\begin{equation}
        \rho_{\vec n|z}= {1\over p(\vec n|z)}
        \sum_\mu A_{\vec n\mu}\rho_z A_{\vec n\mu}^{\dagger} .
\end{equation}
The contribution of this outcome to the average state survival is therefore
\begin{equation}
        p(\vec n|z)\bra{\psi_z}\rho_{\vec n|z}\ket{\psi_z}
        =\sum_\mu
        \left|\bra{\psi_z}A_{\vec n\mu}\ket{\psi_z}\right|^2 .
\end{equation}
Integrating over the outcomes gives $F$ in Eq.~(\ref{eq:Ddef}).
This derivation also shows why the disturbance is sensitive not only to the
POVM elements $A_{\vec n\mu}^\dagger A_{\vec n\mu}$, but to the particular
Kraus realization of the instrument: different postmeasurement states can
have the same statistics and different operation fidelities.
For pure input states this is the average overlap between the original
state and the conditional postmeasurement state, weighted by the outcome
probability.  We refer to Eq.~(\ref{eq:Ddef}) as a full-state operation
fidelity to distinguish it from marginal disturbance functionals that
probe only part of the output system; see Appendix~\ref{app:zhang_relation}.
With this convention, the identity operation gives $D=0$ and an
uninformative guess gives $I=1/2$.

The two figures of merit have a direct but distinct operational meaning.
Since
\begin{equation}
        2I-1=\left\langle \vec n\cdot\vec m\right\rangle
        =\langle\cos\theta\rangle ,
        \label{eq:physical_mean_alignment}
\end{equation}
$I$ quantifies the mean alignment of the guessed and true directions: $I=1/2$
corresponds to a direction chosen independently of the signal, while $I=1$
would correspond to perfect directional inference.  By contrast, $F=1-D$
is the average probability that the conditional output passes a projection
test onto the complete input signal.  It therefore probes preservation of
both local spin states and their coherence across angular-momentum sectors;
$D$ is not an angular estimation error.  The optimal curve $D_{\min}(I)$
thus answers the operational question: what is the smallest loss of
full-state survival that must accompany a prescribed mean directional
accuracy?  Moving along the boundary toward larger $I$ sharpens the
classical directional estimate while reducing the best achievable survival
of the quantum signal.  The antiparallel advantage found below means that
coherence between inequivalent irreducible sectors permits this directional
information to be extracted with a smaller loss of full-state fidelity than
for the irreducible parallel encoding.

It is also useful to distinguish the present score tradeoff from
general information-gain--operation-fidelity or information-balance
relations
\cite{BanaszekPRL,BanaszekDevetak,LeeInfoBalance,HongInfoTradeoff}.
Those relations typically quantify information by an entropy, a mutual
information, or a state-estimation fidelity defined for a general
ensemble.  Here $I$ is instead the task-specific directional score in
Eq.~(\ref{eq:Idef}).  Consequently, a general information-gain bound
cannot be converted into the present $I$--$D$ curve by a universal
change of variables.  Rather, the optimization carried out here
provides the tight, ensemble-specific analogue of such bounds for a
uniformly distributed direction and the score $s(\vec n,\vec
m)$. Although general information-balance relations constrain the
performance of quantum instruments, their information measures do not
generally coincide with the directional score used here. They
therefore do not directly yield the present \(I\)--\(D\) curve. The
covariant optimization performed below instead provides the tight
tradeoff for this specific direction-estimation task.

\subsection{Covariant reduction}
The use of covariance is standard in quantum statistical decision theory
and in the theory of covariant instruments
\cite{Holevo,DaviesLewis,DaviesOpen,OzawaCont,HolevoImprimitivity,HolevoCovInstr,CarmeliCovInstr}.
Suppose that an arbitrary instrument $\{\E_r\}$ is used with a guess rule
$r\mapsto g_r$.  Define the covariantized instrument
\begin{equation}
\widetilde{\E}_h(\rho)=\sum_r
        V_hV_{g_r}^{\dagger}
        \E_r\!\left(V_{g_r}V_h^{\dagger}\rho
        V_hV_{g_r}^{\dagger}\right)
        V_{g_r}V_h^{\dagger} .
        \label{eq:covariantized}
\end{equation}
The equality of the averaged figures of merit before and after
covariantization follows directly from invariance.  For example, the
information of the original strategy can be written as
\begin{equation}
        I=\int dg\sum_r s(g_r,g)
        \Tr\!\left[\E_r(V_g\rho_0V_g^\dagger)\right] .
\end{equation}
Inserting Eq.~(\ref{eq:covariantized}), changing variables by left and right
Haar invariance, and using $s(kg_r,kg)=s(g_r,g)$ leaves this integral
unchanged.  The same calculation applies to the operation fidelity because
both the input projector and the output projector are rotated by the same
unitary.  Hence covariance is not an ansatz: it is a consequence of group
averaging together with linearity of the figures of merit.  Because both
$I$ and $F$ are linear in the instrument before the final optimization,
the group average of an arbitrary strategy preserves the averaged values;
no convexity or extremality assumption is involved at this stage.
For a continuous outcome set the sum is replaced by the corresponding
integral.  Thus no optimality is lost by restricting the search to covariant
instruments.

A covariant instrument satisfies
\begin{equation}
        \E_h(V_g\rho V_g^\dagger)
        =V_g\E_{g^{-1}h}(\rho)V_g^\dagger .
        \label{eq:covariance}
\end{equation}
Equivalently, all maps are generated from a single seed map $\E_0$,
\begin{equation}
        \E_g(\rho)=V_g\E_0(V_g^\dagger\rho V_g)V_g^\dagger .
        \label{eq:covariant_seed_map}
\end{equation}
For direction estimation the physical outcome is not a full rotation
but a point on the sphere, $S^2\simeq G/H$, where the stabilizer
subgroup $H\simeq\mathrm{U}(1)$ consists of rotations around the
fiducial direction $\vec z$.  If $g$ and $gh$, with $h\in H$, send
$\vec z$ to the same direction, they correspond to the same guess.
Moreover, the score depends only on the coset $gH$.

The fiducial signal states used below are invariant under the stabilizer
at the level relevant for the figures of merit.  For parallel spin-coherent
states, rotations around $\vec z$ produce only an overall phase.  For the
antiparallel fiducial state $\ket{j,j}\ket{j,-j}$, the two phases produced by a
rotation around $\vec z$ are opposite and cancel.  Thus the projector
$\proj{\psi_z}$ is invariant under $H$.  The seed instrument may therefore be
averaged over the stabilizer without changing $I$ or $F$.  With the normalized
decomposition $dg=\dn\,dh$, this removes the redundant rotation around the
guessed direction and leaves a covariant instrument labeled only by
$\vec n\in S^2$.

\subsection{Jamio\l{}kowski representation and SDP form}
We use the Jamio\l{}kowski representation of completely positive maps
\cite{Jamio}.  Together with the covariance reduction above, this gives
the standard finite-dimensional form of a covariant-instrument
optimization problem
\cite{DaviesLewis,DaviesOpen,HolevoCovInstr,CarmeliCovInstr}.  For a
completely positive seed map $\E_0$ we define
\begin{equation}
        R_0=(I_{\rm in}\otimes \E_0)\ket{\Phi}\bra{\Phi},
        \qquad
        \ket{\Phi}=\sum_i\ket{i}\ket{i},
        \label{eq:jamio_state}
\end{equation}
so that
\begin{equation}
        \E_0(\rho)=\Tr_{\rm in}
        \left[(\rho^\tau\otimes I_{\rm out})R_0\right],
        \qquad R_0\geq0 .
        \label{eq:choi}
\end{equation}
Here $\tau$ denotes transposition in the basis used to define the
unnormalized maximally entangled vector $\ket{\Phi}$.  We also use the
double-ket notation for vectorization \cite{lop}.  If
$X=\sum_{ij}X_{ij}\ket{i}_{\rm out}\bra{j}_{\rm in}$ is a linear
operator from the input Hilbert space to the output Hilbert space,
then
\begin{equation}
        \dvec{X}=\sum_{ij}X_{ij}\ket{j}_{\rm in}\ket{i}_{\rm out},
        \qquad
        \dbra{X}Y\otimes Z\dvec{X}
        =\Tr\!\left[X^\dagger Z X Y^\tau\right] .
        \label{eq:vectorization_convention}
\end{equation}
With this convention a rank-one Choi operator $R=\dvec{X}\dbra{X}$
represents the single-Kraus map $E (\rho)=X\rho X^\dagger$. The Choi
operator for outcome $g$ is
\begin{equation}
        R_g=(V_g^*\otimes V_g)R_0(V_g^\tau\otimes V_g^\dagger),
        \label{eq:choi_covariant}
\end{equation}
and trace preservation of the nonselective operation is equivalent to
\begin{equation}
        \int dg\,V_g^*\Tr_{\rm out}[R_0]V_g^\tau=I_{\rm in} .
        \label{eq:tp_covariant}
\end{equation}
To obtain the trace constraint explicitly, insert Eq.~(\ref{eq:choi}) in the
condition that the nonselective map is trace preserving.  For the seed one has
\begin{equation}
        \Tr\!\left[\E_0(\rho)\right]
        =\Tr\!\left[(\rho^\tau\otimes I_{\rm out})R_0\right]
        =\Tr\!\left[\rho^\tau\Tr_{\rm out}R_0\right] .\label{eq:sd}
\end{equation}

For the covariant family one averages the input-side operator
$Y=\Tr_{\rm out}R_0$ over the representation $V_g^*$,
which is the action induced by the transposition convention in the Choi
representation, giving Eq.~(\ref{eq:tp_covariant}).  In general the input representation may
contain multiplicities \cite{Holevo,chir1,CarmeliCovInstr}:
\begin{equation}
        V_g^*\simeq\bigoplus_\alpha
        \left(U_g^{(\alpha)}\otimes I_{m_\alpha}\right),
\end{equation}
where $U_g^{(\alpha)}$ acts on the irrep space and $I_{m_\alpha}$ is the
identity on the multiplicity space.  Schur's lemma then averages over the
irrep factor and leaves an operator on the multiplicity factor.  In the
representations treated explicitly in this work, namely $0\oplus1$ and
$\bigoplus_{\ell=0}^{2j}\ell$, each irrep occurs once, so $m_\alpha=1$ and the
condition reduces to one scalar trace constraint per sector.  Equivalent
irreducible components would lead instead to matrix-valued constraints on the
multiplicity spaces and cannot be ignored in general \cite{nota}.
The two quantities of interest are affine-linear functions of the seed:
\begin{equation}
        I=\Tr(R_I R_0),
        \qquad
        D=1-\Tr(R_F R_0).
        \label{eq:linearID}
\end{equation}
The positive operators $R_I$ and $R_F $ are obtained by the group
averages
\begin{align}
        R_I
        &=\int dg\,s(g)
        \left(V_g\ket{\psi_0}\bra{\psi_0}V_g^\dagger\right)^\tau
        \otimes I_{\rm out} .
        \label{eq:RI_general}\\
        R_F 
        &=\int dg\,
        \left(V_g\ket{\psi_0}\bra{\psi_0}V_g^\dagger\right)^\tau
        \otimes
        V_g\ket{\psi_0}\bra{\psi_0}V_g^\dagger,
        \label{eq:Rop_general}
\end{align}
The operators $R_I$ and $R_F $ are obtained by moving all dependence on
the seed map into $R_0$.  For instance, with
$\rho_g=V_g\proj{\psi_0}V_g^\dagger$,
\begin{equation}
        \Tr\!\left[\E_g(\rho_g)\right]
        =\Tr\!\left[(\rho_g^\tau\otimes I)R_g\right]
\end{equation}
and covariance allows the group action to be shifted from $R_g$ to the fixed
seed.  Averaging the resulting coefficient of $R_0$ gives
Eq.~(\ref{eq:RI_general}).  Similarly,
\begin{equation}
        \bra{\psi_g}\E_g(\rho_g)\ket{\psi_g}
        =\Tr\!\left[(\rho_g^\tau\otimes \rho_g)R_g\right],
\end{equation}
which gives Eq.~(\ref{eq:Rop_general}) after the same change of variables.
Thus the primal optimization is linear in the Choi seed.
Here $s(g)$ is the directional score associated with the direction
obtained by applying $g$ to the fiducial direction.  Thus, at fixed
information, the minimum-disturbance problem is the SDP
\begin{align}
\hbox{minimize}\quad &1-\Tr(R_F R_0)\nonumber\\
\hbox{subject to}\quad &R_0\geq0,\nonumber\\
&\Tr(R_I R_0)=I,\nonumber\\
&\int dg\,V_g^*\Tr_{\rm out}[R_0]V_g^\tau=I_{\rm in} .
        \label{eq:general_primal_ID}
\end{align}
The following sections solve this program explicitly for the two
spin-one-half encodings and derive closed endpoint formulas for
arbitrary spin.

\section{Two antiparallel spin-one-half particles}
\label{sec:anti_half}

\subsection{Representation and signal states}
The physical antiparallel pair is
\begin{equation}
        \ket{\psi_{\vec n}}=\ket{\vec n}\ket{-\vec n}
        =(U_g\otimes U_g)\ket{\uparrow}\ket{\downarrow},
        \qquad \vec n=g\vec z .
        \label{eq:physical_antiparallel_half}
\end{equation}
The two-spin Hilbert space decomposes as 
\begin{equation}
        {1\over2}\otimes {1\over2}\simeq0\oplus1 .
        \label{eq:half_decomp}
\end{equation}
We introduce the singlet-triplet Cartesian basis
\begin{equation}
        \{\ket0,\ket x,\ket y,\ket z\},
        \label{eq:cartesian_basis}
\end{equation}
where
\begin{equation}
        \ket0={1\over\sqrt2}
        \left(\ket{\uparrow\downarrow}-\ket{\downarrow\uparrow}\right)
\end{equation}
spans the scalar sector, while $\ket x,\ket y,\ket z$ span the triplet sector
and are chosen as a real Cartesian vector basis.  In particular,
\begin{equation}
        \ket z={1\over\sqrt2}
        \left(\ket{\uparrow\downarrow}+\ket{\downarrow\uparrow}\right),
\end{equation}
with phases of $\ket x$ and $\ket y$ chosen so that
\begin{equation}
        (U_g\otimes U_g)\ket z
        =n_x\ket x+n_y\ket y+n_z\ket z .
\end{equation}
Since
\begin{equation}
        \ket{\uparrow}\ket{\downarrow}
        ={1\over\sqrt2}(\ket0+\ket z),
\end{equation}
the antiparallel signal associated with the unit vector $\vec n$ is
\begin{equation}
        \ket{\psi_{\vec n}}={1\over\sqrt2}
        \left(\ket0+n_x\ket x+n_y\ket y+n_z\ket z\right),
        \label{eq:psi_n_half}
\end{equation}
and the fiducial state is
\begin{equation}
        \ket{\psi_z}={1\over\sqrt2}(\ket0+\ket z).
        \label{eq:psi_z_half}
\end{equation}
The simple scalar-plus-vector form is therefore obtained directly in the
physical $U_g\otimes U_g$ representation.  It is the singlet-triplet analogue
of the operator identity
$\ket{\vec n}\bra{\vec n}=(I+\vec n\cdot\vec\sigma)/2$, but it is written
entirely in the physical two-spin Hilbert space.  The scalar-vector coherence in
Eq.~(\ref{eq:psi_n_half}) is the resource responsible for the antiparallel
advantage in direction estimation \cite{GisinPopescu}.

We use the normalized invariant measure on the sphere,
\begin{equation}
        \int\dn=1,
        \qquad
        \int\dn\,n_i=0,
        \qquad
        \int\dn\,n_i n_j={\delta_{ij}\over3} .
        \label{eq:spherical_moments_half}
\end{equation}
The optimal covariant direction POVM has density
\begin{equation}
        \Pi_{\vec n}=\ket{\eta_{\vec n}}\bra{\eta_{\vec n}},
        \qquad
        \ket{\eta_{\vec n}}=\ket0+\sqrt3\,(n_x\ket x+n_y\ket y+n_z\ket z).
        \label{eq:eta_povm_half}
\end{equation}
Indeed, Eq.~(\ref{eq:spherical_moments_half}) gives
\begin{equation}
        \int\dn\,\Pi_{\vec n}=P_0+P_1=I_{\rm in},
        \qquad P_0=\proj0,
        \qquad P_1=I_{\rm in}-P_0 .
        \label{eq:povm_norm_half}
\end{equation}
For a true $z$ direction, with $x=n_z$, the likelihood of this POVM is
\begin{equation}
        p_{\rm opt}(x)=\left|\langle\eta_{\vec n}|\psi_z\rangle\right|^2
        ={1\over2}(1+\sqrt3 x)^2 .
        \label{eq:popt_half}
\end{equation}
With the directional score $s(x)=(1+x)/2$, this gives the
Gisin-Popescu value \cite{GisinPopescu}
\begin{equation}
        I_{\max}=
        \int_{-1}^{1}{dx\over2}{1+x\over2}{1\over2}(1+\sqrt3x)^2
        ={3+\sqrt3\over6} .
        \label{eq:Imax_half}
\end{equation}

\subsection{Seed of the covariant instrument}
As explained after Eq.~(\ref{eq:sd}) trace preservation of the nonselective operation gives one constraint per
irreducible sector. More
explicitly, if $Y=\Tr_{\rm out}R_0$, Eq.~(\ref{eq:tp_covariant}) gives
\begin{equation}
        \int dg\,V_g^*YV_g^\tau
        ={\Tr(P_0YP_0)\over1}P_0
        +{\Tr(P_1YP_1)\over3}P_1 .
\end{equation}
Equating this expression to $P_0+P_1=I_{\rm in}$ yields
\begin{equation}
        \Tr\!\left[P_0\Tr_{\rm out}(R_0)P_0\right]=1,
        \qquad
        \Tr\!\left[P_1\Tr_{\rm out}(R_0)P_1\right]=3 .
        \label{eq:trace_constraints_half}
\end{equation}
These numbers are simply the dimensions of the irreducible sectors.  Equivalently, with
\begin{equation}
        A_0=P_0\otimes I_4,
        \qquad
        A_1=P_1\otimes I_4,
        \label{eq:A0A1}
\end{equation}
we write
\begin{equation}
        \Tr(A_0R_0)=1,
        \qquad
        \Tr(A_1R_0)=3 .
        \label{eq:trace_A_half}
\end{equation}

The information and operation fidelity are linear functions of $R_0$:
\begin{equation}
        I=\Tr(R_I R_0),
        \qquad
        F=\Tr(R_F R_0),
        \qquad
        D=1-F .
        \label{eq:linear_figures}
\end{equation}
In the ordered basis of Eq.~(\ref{eq:cartesian_basis}) on the input
factor $R_I=B_I\otimes I_4$,  where
\begin{equation}
        B_I=\begin{pmatrix}
        {1\over4}&0&0&{1\over12}\\
        0&{1\over12}&0&0\\
        0&0&{1\over12}&0\\
        {1\over12}&0&0&{1\over12}
        \end{pmatrix} .
        \label{eq:BI_half}
\end{equation}
Equation~(\ref{eq:BI_half}) can be checked directly.  Since
$\ket{\psi_{\vec n}}=(\ket0+n_i\ket i)/\sqrt2$ and
$s(n_z)=(1+n_z)/2$, the input coefficient is
\begin{equation}
        B_I={1\over4}\int d\vec n\,(1+n_z)
        (\ket0+n_i\ket i)(\bra0+n_j\bra j).
\end{equation}
The required moments are those in Eq.~(\ref{eq:spherical_moments_half}); for
example, $\langle0|B_I|0\rangle=1/4$,
$\langle0|B_I|z\rangle=(1/4)\int d\vec n\,n_z^2=1/12$, and
$\langle i|B_I|j\rangle=(1/4)\delta_{ij}/3$.  All other entries vanish by
parity or axial symmetry.
The operator $R_F$ is given by 
\begin{equation}
        R_F =\int\dn\,
        \ket{\psi_{\vec n}}\bra{\psi_{\vec n}}^\tau\otimes
        \ket{\psi_{\vec n}}\bra{\psi_{\vec n}} .
        \label{eq:Rop_half}
\end{equation}
Its matrix elements are $(R_F )_{ab,cd}=M_{abcd}/4$, where $a,b,c,d\in\{0,x,y,z\}$ and the nonzero components are
\begin{align}
        M_{0000}&=1,\nonumber\\
        M_{00ij}&=M_{0i0j}=M_{0ij0}=M_{i00j}=M_{i0j0}=M_{ij00}
        ={\delta_{ij}\over3},\nonumber\\
        M_{ijkl}&={\delta_{ij}\delta_{kl}+\delta_{ik}\delta_{jl}
        +\delta_{il}\delta_{jk}\over15} .
        \label{eq:M_half}
\end{align}
Rotational invariance fixes all nonzero components.  Terms containing an
odd number of vector indices vanish by inversion symmetry.  Terms with two
vector indices give $\int d\vec n\,n_i n_j=\delta_{ij}/3$.  Terms with four
vector indices give the unique isotropic rank-four tensor
\begin{equation}
        \int d\vec n\, n_i n_j n_k n_l
        =a(\delta_{ij}\delta_{kl}+\delta_{ik}\delta_{jl}
        +\delta_{il}\delta_{jk}).
\end{equation}
Contracting $i=j$ and $k=l$ yields $1=\int d\vec n\,(\vec n\cdot\vec
n)^2=15 a$, hence $a=1/15$. Thus $M_{abcd}$ is the component form of the
group average in Eq.~(\ref{eq:Rop_half}), turning the abstract
covariant expression for $R_F$ into an explicit finite matrix for the
SDP.

\section{Optimal tradeoff from the covariant dual problem}
\label{sec:optimal_method}

To trace the boundary of the achievable region we maximize a supporting
linear functional of the operation fidelity and the information.  For a fixed
real parameter $\mu$ this amounts to maximizing
\begin{equation}
        F+\mu I=\Tr[(R_F +\mu R_I)R_0] .
        \label{eq:linear_combination}
\end{equation}
Varying $\mu$ gives the supporting lines of the concave optimal curve
$F(I)$, or equivalently of the convex disturbance curve $D(I)=1-F(I)$.
For an irreducible input representation there is only one trace constraint,
and the optimization reduces to a largest-eigenvalue problem.  In the
antiparallel spin-$1/2$ case the input representation is $0\oplus1$, and the
two sector normalizations lead to
\begin{align}
\hbox{minimize}\quad &\lambda_0+3\lambda_1 \nonumber\\
\hbox{subject to}\quad
&K_\mu(\lambda_0,\lambda_1)
=\lambda_0A_0+
\lambda_1A_1-(R_F +\mu R_I)\ge0 .
        \label{eq:dual_mu}
\end{align}
This dual follows from the Lagrangian of the primal maximization of
$\Tr[(R_F+\mu R_I)R_0]$ under
$\Tr(A_0R_0)=1$, $\Tr(A_1R_0)=3$, and $R_0\ge0$:
\begin{equation}
        {\cal L}(R_0,\lambda_0,\lambda_1)
        =\lambda_0+3\lambda_1
        -\Tr\!\left[\left(\lambda_0A_0+
        \lambda_1A_1-R_F -\mu R_I\right)R_0\right].
\end{equation}
The Lagrangian is bounded above over $R_0\ge0$ iff the operator in
parentheses is positive.  At optimality, complementary slackness gives
\cite{bv}
\begin{equation}
        K_\mu(\lambda_0,\lambda_1)R_0^\star=0,
\end{equation}
so the support of every optimal Choi seed is contained in the zero eigenspace
of the optimal dual slack.

For the antiparallel spin-$1/2$ problem the active kernel of the dual
slack is one-dimensional along the generic interior branch.  Let $\dvec{X_\mu}$ be the corresponding
normalized kernel vector.  The sector constraints read
\begin{equation}
        \dbra{X_\mu}A_0\dvec{X_\mu}={1\over4},
        \qquad
        \dbra{X_\mu}A_1\dvec{X_\mu}={3\over4},
        \label{eq:X_mu_norm}
\end{equation}
and the optimal seed is
\begin{equation}
        R_0(\mu)=4\,\dvec{X_\mu}\dbra{X_\mu} .
        \label{eq:R0_mu_rankone}
\end{equation}
Appendix~\ref{app:rank_one_simple_kernel} gives the general reason for
this normalization: when the active kernel of the dual
slack is one-dimensional, the KKT
conditions force its sector weights to be proportional to the sector
dimensions.

This rank-one construction describes the generic exposed branch.  At endpoints
or isolated values of $\mu$ the active kernel may be degenerate.  If
$\ker K_\mu={\rm span}\{\dvec{X_1},\ldots,\dvec{X_d}\}$, complementary
slackness gives only ${\rm supp}\,R_0^\star\subseteq\ker K_\mu$.  The most
general optimal seed on that exposed face is
\begin{equation}
        R_0^\star=\sum_{r,s=1}^d G_{rs}\,
        \dvec{X_r}\dbra{X_s},
        \qquad G\ge0,
        \label{eq:degenerate_kernel_seed}
\end{equation}
where $G$ is chosen to satisfy
\begin{equation}
        \sum_{r,s}G_{rs}\dbra{X_s}A_0\dvec{X_r}=1,
        \qquad
        \sum_{r,s}G_{rs}\dbra{X_s}A_1\dvec{X_r}=3 .
        \label{eq:degenerate_kernel_constraints}
\end{equation}
Thus a degenerate active kernel leaves only a reduced finite-dimensional SDP
inside the exposed face; a diagonal mixture is sufficient only when the sector
normalizations can be met without coherences among the kernel vectors.

In the simple-kernel case,
\begin{equation}
        I(\mu)=4\dbra{X_\mu}R_I\dvec{X_\mu},
        \qquad
        F(\mu)=4\dbra{X_\mu}R_F \dvec{X_\mu},
        \qquad
        D(\mu)=1-F(\mu).
        \label{eq:IF_mu}
\end{equation}
Axial symmetry around the fiducial direction allows the seed Kraus operator to be chosen real and of the form
\begin{equation}
        X_\mu=
        \begin{pmatrix}
        u_\mu&0&0&v_\mu\\
        0&w_\mu&0&0\\
        0&0&w_\mu&0\\
        r_\mu&0&0&s_\mu
        \end{pmatrix} .
        \label{eq:X_mu_form}
\end{equation}
The corresponding covariant instrument is
\begin{equation}
        \E_{\vec n}^{(\mu)}(\rho)=
        A_{\vec n}^{(\mu)}\rho A_{\vec n}^{(\mu)\dagger},
        \qquad
        A_{\vec n}^{(\mu)}=2 V_g X_\mu V_g^\dagger,
        \qquad \vec n=g\vec z .
        \label{eq:optimal_filter_half}
\end{equation}
The scalar-vector mixing terms $v_\mu$ and $r_\mu$ are the operational
signature of the coherence between inequivalent irreducible sectors.  They
also show why the optimal curve is not, in general, obtained by convexly
interpolating between the identity channel and a measure-and-reprepare map.

For every feasible $R_0$,
\begin{equation}
        \Tr[(R_F +\mu R_I)R_0]
        \le \lambda_0(\mu)+3\lambda_1(\mu),
        \label{eq:dual_bound_mu}
\end{equation}
and equality is attained by Eq.~(\ref{eq:R0_mu_rankone}) in the simple-kernel
case, or by any feasible seed of Eq.~(\ref{eq:degenerate_kernel_seed}) in a
degenerate case.  Hence Eq.~(\ref{eq:IF_mu}) generates the globally optimal
covariant tradeoff.

\section{Endpoints and comparison with measure-and-reprepare}
\label{sec:endpoints}

At $\mu=0$ the linear objective is just $F$.  Trivially, the optimal instrument is the
identity channel and 
\begin{equation}
        I(0)={1\over2},
        \qquad
        F(0)=1,
        \qquad
        D(0)=0 .
        \label{eq:identity_endpoint}
\end{equation}
The opposite endpoint is obtained as $\mu\to\infty$, where the information is
maximized first.  The optimal POVM is Eq.~(\ref{eq:eta_povm_half}), but the
output state need not be the guessed signal state.  At this endpoint
the seed Kraus operator has the rank-one form
\begin{equation}
  X_{\infty}
  ={1\over2}\ket{\varphi_z}\bra{\eta_z},
        \qquad
        \ket{\varphi_z}=\cos\beta\,\ket0+
        \sin\beta\,\ket z .
        \label{eq:endpoint_seed}
\end{equation}
The operation fidelity is obtained by applying the fixed-POVM output
optimization of Appendix~\ref{app:fixed_povm_Q} to the optimal POVM.
The maximum-information likelihood is
\begin{equation}
        |\langle\eta_{\vec n}|\psi_z\rangle|^2
        ={1\over2}(1+\sqrt3 n_z)^2 .
\end{equation}
Thus the remaining output-state optimization is governed by
\begin{equation}
        Q=\int d\vec n\,
        |\langle\eta_{\vec n}|\psi_z\rangle|^2
        V_g^\dagger\ket{\psi_z}\bra{\psi_z}V_g,
        \qquad \vec n=g\vec z .
        \label{eq:Q_endpoint_integral_half}
\end{equation}
Since the likelihood and the fiducial input are invariant under rotations
around $z$, $Q$ commutes with the stabilizer of $z$.  Hence the relevant
maximizing output vector belongs to $\mathrm{span}\{\ket0,\ket z\}$, and the
transverse states $\ket x,\ket y$ do not contribute to the largest eigenvalue.
Direct evaluation of the spherical moments gives
\begin{equation}
        Q\big|_{\{0,z\}}=
        \begin{pmatrix}
        {1\over2}&{\sqrt3\over6}\\[2mm]
        {\sqrt3\over6}&{7\over30}
        \end{pmatrix} .
        \label{eq:Q_endpoint}
\end{equation}
The diagonal entries are the likelihood-weighted survival contributions of the
scalar and longitudinal output components, while the off-diagonal entry is the
corresponding scalar-longitudinal coherence.  The largest eigenvector fixes
the best fiducial output state $\ket{\varphi_z}$, and the largest eigenvalue is
the endpoint operation fidelity.  The eigenvalues are
\begin{equation}
        q_\pm={11\pm\sqrt{91}\over30},
\end{equation}
so the maximum operation fidelity, i.e. the largest eigenvalue of $Q$, is
\begin{equation}
        F(I_{\max})={11+\sqrt{91}\over30},
        \label{eq:F_endpoint_opt}
\end{equation}
and therefore the minimal disturbance is
\begin{equation}
        D(I_{\max})=1-\lambda_{\max}(Q)
        ={19-\sqrt{91}\over30}
        \simeq0.31535 .
        \label{eq:D_endpoint_opt}
\end{equation}
The angle $\beta$ in Eq.~(\ref{eq:endpoint_seed}) is fixed by the
normalized eigenvector of $Q$ associated with this largest eigenvalue.
This shows explicitly that the optimal output is a covariant family
generated from a state in the scalar-longitudinal subspace, rather
than being imposed a priori to coincide with the guessed signal state.
The disturbance in Eq.~(\ref{eq:D_endpoint_opt}) is smaller than the
disturbance of the measure-and-reprepare endpoint, where the output
state is forced to be the guessed signal state $\ket{\psi_{\vec n}}$.  In that case
\begin{equation}
        F_{\rm mr}(I_{\max})
        =
        \int_{-1}^{1}{dx\over2}\,
        {1\over2}(1+\sqrt3 x)^2
        \left({1+x\over2}\right)^2
        =
        {11+\sqrt{75}\over30},
\end{equation}
and therefore
\begin{equation}
        D_{\rm mr}(I_{\max})
        =
        {19-\sqrt{75}\over30}
        \simeq0.34466 .
\end{equation}
Thus, even at maximal information, minimum disturbance requires optimizing the
postmeasurement state, not merely repreparing the guessed signal
state.  A similar result appeared also at the endpoint of the optimal
information-disturbance tradeoff in quantum-state discrimination \cite{BuscemiSacchi}.

\section{Parallel-spin benchmark}
\label{sec:parallel_benchmark}

The parallel two-spin benchmark is useful because it is the irreducible version
of the same optimization problem.  Two parallel spin-one-half particles occupy
the symmetric subspace and are equivalent to a single spin coherent state with
total spin $J=1$.  We denote the coherent state in this subspace by
\begin{equation}
        \ket{\phi_{\vec n}}=U_g^{(J)}\ket{J,J},
        \qquad \vec n=g\vec z .
        \label{eq:parallel_phi_n}
\end{equation}
The representation is irreducible.  Therefore, unlike the antiparallel case,
there is a single trace-preserving constraint for the covariant Choi seed,
\begin{equation}
        \Tr R_0=2J+1=3 .
        \label{eq:parallel_trace_constraint}
\end{equation}
This irreducibility makes the benchmark technically simpler: optimizing a
linear combination of information and operation fidelity reduces to a
maximum-eigenvalue problem rather than to a semidefinite problem with several
sector multipliers.

For the directional score used throughout this paper, the two linear operators
are
\begin{align}
        R_I^{\parallel}
        &=\int\dn\,{1+n_z\over2}
        \ket{\phi_{\vec n}}\bra{\phi_{\vec n}}^{\tau}\otimes I_{J=1},
        \label{eq:RI_parallel}\\
        R_F ^{\parallel}
        &=\int\dn\,
        \ket{\phi_{\vec n}}\bra{\phi_{\vec n}}^{\tau}
        \otimes
        \ket{\phi_{\vec n}}\bra{\phi_{\vec n}} .
        \label{eq:Rop_parallel}
\end{align}
For each real parameter $\mu$ one defines
\begin{equation}
        C_{\mu}^{\parallel}=R_F ^{\parallel}+
        \mu R_I^{\parallel} .
        \label{eq:Cmu_parallel}
\end{equation}
Since the normalization constraint is only $\Tr R_0=3$, the dual slack is
\begin{equation}
        K_{\mu}^{\parallel}=\lambda I_3\otimes I_3-C_{\mu}^{\parallel},
        \label{eq:Kmu_parallel}
\end{equation}
and the smallest feasible $\lambda$ is the largest eigenvalue of
$C_{\mu}^{\parallel}$.  Hence, if
\begin{equation}
        C_{\mu}^{\parallel}\dvec{\chi_\mu}=c_{\max}(\mu)\dvec{\chi_\mu},
        \qquad \dbra{\chi_\mu}|\chi_\mu \rangle \!\rangle =1,
        \label{eq:parallel_eigenproblem}
\end{equation}
then the optimal Choi seed is
\begin{equation}
        R_0^{\parallel}(\mu)=3\dvec{\chi_\mu}\dbra{\chi_\mu} .
        \label{eq:parallel_R0_mu}
\end{equation}
The corresponding optimal benchmark curve is
\begin{equation}
        I^{\parallel}(\mu)=3
        \dbra{\chi_\mu}R_I^{\parallel}\dvec{\chi_\mu},
        \qquad
        D^{\parallel}(\mu)=1-3
        \dbra{\chi_\mu}R_F ^{\parallel}\dvec{\chi_\mu} .
        \label{eq:parallel_ID_mu}
\end{equation}
Equivalently, writing the eigenvector in vectorized form as the Kraus seed
$\chi_\mu$, the optimal covariant instrument has the form
\begin{equation}
        \E_g^{\parallel}(\rho)=3
        U_g^{(J)}\chi_\mu U_g^{(J)\dagger}\,\rho\,
        U_g^{(J)}\chi_\mu^{\dagger} U_g^{(J)\dagger} .
        \label{eq:parallel_filter_benchmark}
\end{equation}
This is the irreducible counterpart of the dual-kernel construction used for
antiparallel spins.  In the antiparallel case the scalar and vector sectors
force two multipliers $\lambda_0,\lambda_1$; in the parallel case the single
multiplier $\lambda$ is just the largest eigenvalue of
$C_{\mu}^{\parallel}$.

\begin{figure}[t]
\begin{center}
\includegraphics[width=0.48\textwidth]{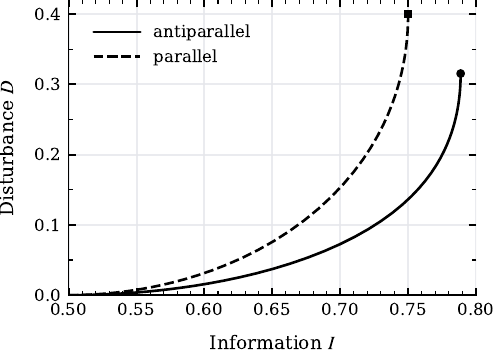}
\caption{Optimal information-disturbance tradeoffs for
  two-spin direction encoding.  The solid curve is the antiparallel
  spin-$1/2$ case obtained from the two-sector dual-kernel
  construction.  The dashed curve is the parallel-spin benchmark,
  equivalent to the irreducible spin-$J=1$ spin-coherent-state
  problem. The markers denote the fully informative endpoints:
  $P_{\rm anti}=((3+\sqrt3)/6,(19-\sqrt{91})/30)$ and
  $P_{\parallel}=(3/4,2/5)$, where each ordered pair is $(I,D)$.}  
\label{fig:benchmark_tradeoff_comparison_PRA}
\end{center}
\end{figure}

At the fully informative endpoint the coherent-state POVM in the spin-$J$
subspace has density \cite{Perelomov,Arecchi,SacchiSCS}
\begin{equation}
        \Pi_{\vec n}^{\parallel}=(2J+1)
        \ket{\phi_{\vec n}}\bra{\phi_{\vec n}} .
        \label{eq:parallel_endpoint_povm}
\end{equation}
For a true $z$ direction,
\begin{equation}
        \left|\langle\phi_{\vec n}|\phi_z\rangle\right|^2
        =\left({1+x\over2}\right)^{2J},
        \qquad x=n_z,
        \label{eq:parallel_overlap_J}
\end{equation}
and hence, for $J=1$, the endpoint information is 
\begin{equation}
        I^{\parallel}_{\max}
        =\int_{-1}^{1}{dx\over2}{1+x\over2}
        3\left({1+x\over2}\right)^2
        ={3\over4} .
        \label{eq:Iparallel_max_benchmark}
\end{equation}
At the same endpoint, the POVM is fixed but the output state is not.  Using
Appendix~\ref{app:fixed_povm_Q}, the residual optimization is described by
\begin{equation}
        Q^{\parallel}=3\int\dn\,
        \left|\langle\phi_{\vec n}|\phi_z\rangle\right|^2
        U_g^{(J)\dagger}\ket{\phi_z}\bra{\phi_z}U_g^{(J)} .
        \label{eq:Qparallel_endpoint}
\end{equation}
For any fiducial output $\ket{\varphi_z}$ in the spin-$J$ space,
\begin{equation}
  F(\varphi)
        =\bra{\varphi_z}Q^{\parallel}\ket{\varphi_z}.
\end{equation}
Axial symmetry makes $Q^{\parallel}$ diagonal in the $J_z$ basis: the weight
$|\langle\phi_{\vec n}|\phi_z\rangle|^2$ depends only on $n_z$, so the azimuthal
integral removes all matrix elements with different magnetic quantum numbers.
For $J=1$ its eigenvalues are
\begin{equation}
        q_{1}={3\over5},
        \qquad
        q_{0}={3\over10},
        \qquad
        q_{-1}={1\over10} .
        \label{eq:Qparallel_eigenvalues}
\end{equation}
Thus the maximum is obtained by repreparing the guessed coherent state itself,
and
\begin{equation}
  F(I^{\parallel}_{\max })
  ={3\over5},
        \qquad
        D(I^{\parallel}_{\max })        =1-{3\over5}={2\over5} .
        \label{eq:Dparallel_endpoint_benchmark}
\end{equation}
The endpoint comparison with the optimized antiparallel construction is then
\begin{equation}
        I^{\rm anti}_{\max}={3+\sqrt3\over6}>{3\over4}=I^{\parallel}_{\max},
        \label{eq:endpoint_info_comparison_benchmark}
\end{equation}
while
\begin{equation}
        D(I^{\rm anti}_{\max })={19-\sqrt{91}\over30}
        <{2\over5}=D(I^{\parallel}_{\max }) .
        \label{eq:endpoint_disturbance_comparison_benchmark}
\end{equation}
Thus the antiparallel encoding remains advantageous even with respect
to the figure of merit of disturbance.  Physically, the scalar-vector
coherence of the antiparallel signal supplies the directional
dependence of the measurement statistics while allowing the
conditional output to retain a larger overlap with the complete input
state.  The comparison is therefore not merely that antiparallel spins
are easier to estimate: at equal extracted score they also admit a
less invasive optimal instrument. The complete comparison of the
optimal information-disturbance tradeoffs for the two spin-$1/2$ case
is shown in Fig. \ref{fig:benchmark_tradeoff_comparison_PRA}.

\begin{figure}[t]
\begin{center}
\includegraphics[width=0.48\textwidth]{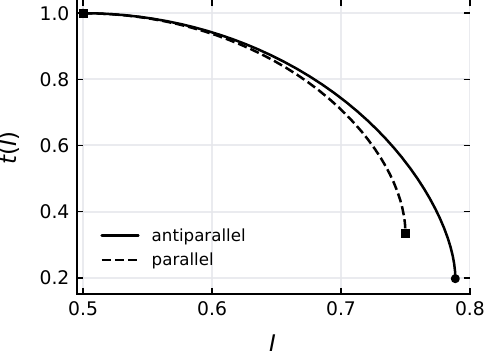}
\caption{Normalized squared traces of the Kraus seeds along the optimal
tradeoffs for two spin-\(1/2\) particles.  The solid curve is the
antiparallel encoding, with \(t(I)=|\operatorname{Tr}X(I)|^2/4\);
the dashed curve is the parallel encoding, with
\(t(I^\parallel)=|\operatorname{Tr}\chi(I^\parallel)|^2/3\).
}
\label{ti}
\end{center}
\end{figure}

A global quantity that in both cases continuously characterizes the
covariant instruments that optimize the information-disturbance
tradeoff is the normalized squared trace of the Kraus seeds $X_\mu$
and $\chi_ \mu$ in Eqs.~(\ref{eq:optimal_filter_half}) and
(\ref{eq:parallel_filter_benchmark}), respectively.  Expressed as a
function of the extracted information, for antiparallel encoding the
scalar $t(I)=|\operatorname{Tr}X(I) |^2/4$ equals $1$ at the identity
endpoint, where no information is extracted beyond the random score
$I=1/2$, and decreases concavely to $
|\operatorname{Tr}X(I_{\max})|^2/4 =(\cos\beta+\sqrt3\sin\beta )^2/16 \simeq0.19707$ at the maximum-information endpoint, where
  $\ket{\varphi_z}=\cos\beta\ket0+\sin\beta\ket z$ is the
  largest-eigenvalue eigenvector of the endpoint operator $Q$ in
  Eq. (\ref{eq:Q_endpoint}). For the parallel case
  $t(I^\parallel )=|\operatorname{Tr}\chi (I^ \parallel ) |^2/3$ equals $1$ at the identity
  endpoint and $1/3$ at the maximum-information value $I_{\max }^\parallel =3/4$.  We
  plot the results of $t(I)$ for  both cases in Fig. \ref{ti}.

\section{Higher-spin antiparallel states}
\label{sec:higher_spin}
Here we extend our study to antiparallel spin-coherent states of
arbitrary spin $j$,
\begin{equation}
        \ket{\psi_{\vec n}^{(j)}}=\ket{\vec n}_j\ket{-\vec n}_j
        =(U_g^{(j)}\otimes U_g^{(j)})\ket{j,j}\ket{j,-j},
        \qquad \vec n=g\vec z .
        \label{eq:higher_signal}
\end{equation}
The representation is the ordinary two-spin representation
\begin{equation}
        V_g^{(j)}=U_g^{(j)}\otimes U_g^{(j)},
\end{equation}
and it decomposes as
\begin{equation}
        j\otimes j\simeq\bigoplus_{\ell=0}^{2j}\ell .
        \label{eq:higher_decomp}
\end{equation}
The fiducial state decomposes as \cite{Edmonds}
\begin{equation}
        \ket{j,j}\ket{j,-j}=\sum_{\ell=0}^{2j}c_\ell\ket{\ell,0},
        \qquad
        c_\ell^2=(2\ell+1){[(2j)!]^2\over(2j-\ell)!(2j+\ell+1)!} .
        \label{eq:higher_c_l}
\end{equation}
The coefficients in Eq.~(\ref{eq:higher_c_l}) are the Clebsch-Gordan
coefficients
\begin{equation}
        c_\ell=\langle j,j;j,-j|\ell,0\rangle,
\end{equation}
up to an overall phase, chosen here so that all $c_\ell$ are nonnegative.
Normalization follows from completeness of the coupled basis,
$\sum_{\ell=0}^{2j}c_\ell^2=1$.  The factorial expression shows that all
sectors allowed by angular-momentum addition appear, so the fiducial
antiparallel signal is not confined to a single irreducible representation.
The optimal POVM seed for direction estimation is
$\ket{\eta_z^{(j)}}\bra{\eta_z^{(j)}}$, with 
\begin{equation}
  \ket{\eta_z^{(j)}}
  =\sum_{\ell=0}^{2j}\sqrt{2\ell+1}\ket{\ell,0} .
        \label{eq:higher_eta}
\end{equation}
The normalization of the POVM seed follows from the orthogonality relation
for irreducible matrix elements.  In fact, in the sector $\ell$, the orbit of
$\ket{\ell,0}$ satisfies
\begin{equation}
        \int d\vec n\,U_g^{(\ell)}\ket{\ell,0}\bra{\ell,0}
        U_g^{(\ell)\dagger}={\Delta _\ell\over2\ell+1},
\end{equation}
where $\Delta _\ell$ denotes the projectors onto the irreducible sectors,
while cross terms between inequivalent irreducible sectors average to
zero.  Therefore the integral of $\ket{\eta_{\vec
    n}^{(j)}}\bra{\eta_{\vec n}^{(j)}}$ is $\sum_\ell \Delta
_\ell=I_{\rm in}$. 
Defining 
\begin{equation}
        A_\ell=\Delta _\ell\otimes I_{\rm out} ,
        \label{eq:higher_A_l}
\end{equation}
the trace constraints are
\begin{equation}
        \Tr(A_\ell R_0)=2\ell+1,
        \qquad \ell=0,1,\ldots,2j .
        \label{eq:higher_constraints}
\end{equation}
The supporting-line optimization is therefore
\begin{align}
\hbox{minimize}\quad &\sum_{\ell=0}^{2j}(2\ell+1)\lambda_\ell
\nonumber\\
\hbox{subject to}\quad
&K_\mu^{(j)}=\sum_{\ell=0}^{2j}\lambda_\ell A_\ell
-(R_F ^{(j)}+
\mu R_I^{(j)})\ge0 .
        \label{eq:higher_dual}
\end{align}
The optimal Choi seed is supported on the kernel of $K_\mu^{(j)}$.
This is the higher-spin analogue of the irreducible
maximum-eigenvector construction, but with $2j+1$ trace multipliers
replacing the single multiplier of an irreducible representation.

Although the dual kernel can be multidimensional, the general
rank-reduction argument of Appendix~\ref{app:rank_bound} shows that
there always exists an optimal covariant Choi seed with
$\operatorname{rank}R_0^\star\le\lfloor\sqrt{2j+1}\rfloor$.  Hence the
$j=1$ antiparallel tradeoff admits a rank-one optimal seed.

\begin{figure}[t]
\begin{center}
\includegraphics[width=0.48\textwidth]{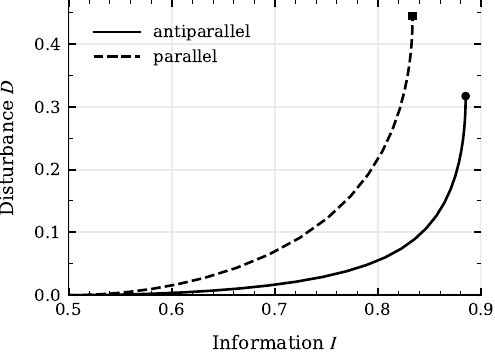}
\caption{Optimal information-disturbance tradeoffs for
  two-spin direction encoding with $j=1$.  The solid curve is the
  antiparallel encoding, obtained from the three-sector dual-kernel
  construction for $1\otimes1\simeq0\oplus1\oplus2$.  The dashed
  curve is the parallel encoding, equivalent to the irreducible
  spin-$J=2$ spin-coherent-state problem. The markers denote the fully informative endpoints,
  labeled $P_{\rm anti}=(0.884773,0.316957)$ and
  $P_{\parallel}=(5/6,4/9)$ in $(I,D)$ coordinates.  }
\label{fig:benchmark_tradeoff_j1_PRA}
\end{center}
\end{figure}

The optimal information-disturbance tradeoff for the case of two
spins $j=1$ is represented in
Fig.~\ref{fig:benchmark_tradeoff_j1_PRA}, where the comparison with
the case of parallel encoding is also reported.

\subsection*{Numerical evaluation of the tradeoff curves}
The curves in Figs.~\ref{fig:benchmark_tradeoff_comparison_PRA} and
\ref{fig:benchmark_tradeoff_j1_PRA} were generated from the
supporting-line parameterization in
Eq.~(\ref{eq:linear_combination}). To resolve both endpoints, we
sampled the compactified parameter $\tau={\mu\over1+\mu} $ uniformly
at $2001$ values $\tau_k=k/2001$, $k=0,\ldots,2000$, corresponding to
$0\leq\mu\leq2000$, and converted each sampled value to
$\mu _k=\tau _k /(1-\tau _k)$. The maximum-information endpoint $\mu\to\infty $
was included separately using the analytical endpoint calculation. At
each value of \(\mu\), the corresponding dual problem was solved: the
two-multiplier problem in Eq.~(\ref{eq:dual_mu}) for antiparallel
spin-\(1/2\), the largest-eigenvalue problem in
Eq.~(\ref{eq:parallel_eigenproblem}) for the irreducible parallel
benchmark, and the three-multiplier specialization of
Eq.~(\ref{eq:higher_dual}) for antiparallel spin \(j=1\). Positivity
was imposed by requiring the minimum eigenvalue of the dual slack
operator to be nonnegative. At the optimum, its null space was
computed numerically. For a one-dimensional null space, the normalized
kernel vector directly gives the rank-one Choi seed through
Eqs.~(\ref{eq:R0_mu_rankone}) and
(\ref{eq:simple_kernel_rankone_seed}). If a degenerate null space is
encountered, the reduced positive feasibility problem in
Eqs.~(\ref{eq:degenerate_kernel_seed}) and
(\ref{eq:degenerate_kernel_constraints}) is solved within that null
space. The values of \(I\) and \(D\) were finally evaluated from the
resulting primal seed. Double-precision arithmetic was used, with the dual
feasibility and zero-eigenvalue residuals required to be below $10^{-10}$. 
\section{Endpoint information and disturbance for arbitrary spin}
\label{sec:endpoints_arbitrary_j}

We now collect the endpoint quantities for two spin-$j$ coherent states,
both for antiparallel and parallel encodings.  The endpoint considered here
is the fully informative endpoint: the covariant POVM is chosen to maximize
the directional information, and, once this POVM is fixed, the output state
is chosen so as to maximize the operation fidelity.  This is important
because the minimum-disturbing endpoint need not coincide with a
measure-and-reprepare channel in which the guessed signal state itself is
reprepared.

\subsection{Antiparallel encoding}
Using the POVM seed of Eq.~(\ref{eq:higher_eta}) for
direction estimation and defining
\begin{equation}
        \alpha_\ell=\sqrt{2\ell+1}\,c_\ell,
        \label{eq:endpoint_alpha_l}
\end{equation}
we can write the maximum-information likelihood for a state with 
relative angle $\theta$, with
$x=n_z=\cos\theta$, as
\begin{equation}
  p_j(x)=
        \left|\langle\eta_{\vec n}^{(j)}|\psi_z^{(j)}\rangle\right|^2
        =T_j(x)^2 ,
        \label{eq:endpoint_antiparallel_likelihood}
\end{equation}
where 
\begin{equation}
        T_j(x)
        =
        \sum_{\ell=0}^{2j}\alpha_\ell P_\ell(x)
        \label{eq:endpoint_Tj}
\end{equation}
is given in terms of the Legendre polynomials $P_\ell(x)$, since the
matrix element of a rotated $m=0$ vector is
$\langle\ell,0|U_g^{(\ell)}|\ell,0\rangle=P_\ell(\cos\theta)$
\cite{Edmonds}.  The maximum information is then
\begin{equation}
        I^{\rm anti,j}_{\max}
        =
        \int_{-1}^{1}{dx\over2}
        {1+x\over2}\,T_j(x)^2 .
        \label{eq:Ianti_endpoint_integral}
\end{equation}
Using the Legendre orthogonality and recurrence relations \cite{pleg}
\begin{align}
        &\int_{-1}^{1}{dx\over2}P_\ell(x)P_{\ell'}(x)
  ={\delta_{\ell\ell'}\over2\ell+1} , \\
  &         xP_\ell(x)
        =
        {\ell+1\over2\ell+1}P_{\ell+1}(x)
        +
        {\ell\over2\ell+1}P_{\ell-1}(x),
        \label{eq:endpoint_legendre_recurrence}
\end{align}
one obtains
\begin{equation}
        I^{\rm anti,j}_{\max}
        =
        {1\over2}
        +
        \sum_{\ell=0}^{2j-1}
        {\ell+1\over\sqrt{(2\ell+1)(2\ell+3)}}
        c_\ell c_{\ell+1}.
        \label{eq:Ianti_max_arbitrary_j}
\end{equation}
Since multiplication by $x$ connects $P_\ell$ only to $P_{\ell+1}$ and
$P_{\ell-1}$, only coherences between adjacent irreducible sectors
enter the information gain above a random guess.  Equivalently,
substituting Eq.~(\ref{eq:higher_c_l}) gives
\begin{equation}
\begin{split}
        I^{\rm anti,j}_{\max}
        =
        {1\over2}
        +
        \sum_{\ell=0}^{2j-1}
        (\ell+1)
        {[(2j)!]^2\over
        \sqrt{
        (2j-\ell)!(2j+\ell+1)!
        (2j-\ell-1)!(2j+\ell+2)!
        }} .
\end{split}
        \label{eq:Ianti_max_factorial_arbitrary_j}
\end{equation}

We now optimize the endpoint disturbance.  At maximum information the POVM is
fixed by Eq.~(\ref{eq:higher_eta}), but the output state after each outcome is
still a variational degree of freedom.  Applying the fixed-POVM construction of
Appendix~\ref{app:fixed_povm_Q}, with likelihood $T_j(n_z)^2$, gives the
positive operator
\begin{equation}
        Q_j
        =
        \int\dn\,
        T_j(n_z)^2\,
        V_g^\dagger
        \ket{\psi_z^{(j)}}\bra{\psi_z^{(j)}}
        V_g ,
        \qquad \vec n=g\vec z .
        \label{eq:Qj_endpoint_definition}
\end{equation}
For a covariant output generated by a fiducial vector $\ket{\varphi_z}$ in
$\bigoplus_{\ell=0}^{2j}{\cal H}_\ell$, the endpoint fidelity is
\begin{equation}
        F^{\rm anti,j}(\varphi)
        =\bra{\varphi_z}Q_j\ket{\varphi_z}.
        \label{eq:Fanti_j_Q_quadratic}
\end{equation}
Thus
\begin{equation}
        F(I^{\rm anti,j}_{\max})
        =
        \lambda_{\max}(Q_j),
        \qquad
        D (I^{\rm anti,j}_{\max})
        =
        1-\lambda_{\max}(Q_j).
        \label{eq:Danti_endpoint_lambda_max}
\end{equation}

Because $T_j(x)^2$ is invariant under rotations around the $z$ axis,
$Q_j$ is block diagonal in the magnetic quantum number $m$.  We denote
by $d^\ell_{m0}(\theta)$ the reduced Wigner rotation matrix element in
the sector of total angular momentum $\ell$.  As shown in
Appendix~\ref{app:Q_antiparallel_SU2}, in the coupled basis
$\ket{\ell,m}$, the block with fixed $m$ has matrix elements
\begin{equation}
        \left[Q_j^{(m)}\right]_{\ell\ell'}
        =
        c_\ell c_{\ell'}
        \int_{-1}^{1}{dx\over2}\,
        T_j(x)^2\,
        d^\ell_{m0}(\theta)\,
        d^{\ell'}_{m0}(\theta),
        \label{eq:Qj_m_block}
\end{equation}
where $x=\cos\theta$ and $\ell,\ell'=|m|,|m|+1,\ldots,2j$. Therefore
\begin{equation}
        F(I^{\rm anti,j}_{\max})
        =
        \max_m \lambda_{\max}\!\left(Q_j^{(m)}\right),
        \qquad
        D(I^{\rm anti,j}_{\max})
        =
        1-
        \max_m \lambda_{\max}\!\left(Q_j^{(m)}\right).
        \label{eq:Danti_endpoint_block_formula}
\end{equation}
The symmetry $m\leftrightarrow -m$ allows one to restrict the maximization to
$m\geq0$.

For $j=1/2$, the matrix in Eq.~(\ref{eq:Qj_m_block}) contains an $m=0$
block and one-dimensional $|m|=1$ blocks; the largest eigenvalue belongs to the
$m=0$ block and gives Eq.~(\ref{eq:Q_endpoint}).  For higher spin, all
numerical calculations reported below give the largest eigenvalue in the
$m=0$ block.  We have not found a general analytic proof that this must hold
for every $j$: although $Q_j$ is positive and block diagonal in $m$, the block
kernels contain reduced Wigner functions $d^\ell_{m0}(\theta)$, whose signs
and relative magnitudes do not imply a simple ordering of the block norms as
$|m|$ increases.  We therefore retain the rigorous maximization over $m$ in
Eq.~(\ref{eq:Danti_endpoint_block_formula}); the statement that $m=0$ is the
maximizing block is reported as numerical evidence for the values in
Table~\ref{tab:endpoint_values_j} up to $j=5$.

\subsection{Parallel encoding}

Two parallel spin-$j$ coherent states are supported in the maximal
total-spin sector and are equivalent to a single spin coherent state
with total spin $J=2j$. From Eqs. (\ref{eq:parallel_endpoint_povm})
and (\ref{eq:parallel_overlap_J}), the likelihood at the fully
informative endpoint is
\begin{equation}
        p_J(x)
        =
        (2J+1)
        \left({1+x\over2}\right)^{2J}.
        \label{eq:endpoint_parallel_likelihood}
\end{equation}
The endpoint information is then 
\begin{equation}
  I^{\parallel,j}_{\max}
        =
        \int_{-1}^{1}{dx\over2}
        {1+x\over2}\,p_J(x)
        =
        {2J+1\over2J+2}
        =
        {4j+1\over4j+2}.
        \label{eq:Iparallel_max_arbitrary_j}
\end{equation}
At the same endpoint, the output state is optimized after the maximum-information
coherent-state POVM has been fixed.  Appendix~\ref{app:fixed_povm_Q} gives
\begin{equation}
        Q_J^{\parallel}
        =(2J+1)\int\dn\,
        \left|\langle\phi_{\vec n}|\phi_z\rangle\right|^2
        U_g^{(J)\dagger}\ket{J,J}\bra{J,J}U_g^{(J)},
        \qquad \vec n=g\vec z .
        \label{eq:Qparallel_arbitrary_J}
\end{equation}
For a fiducial output $\ket{\varphi_z}$,
\begin{equation}
        F_{\max I}^{\parallel,j}(\varphi)
        =\bra{\varphi_z}Q_J^{\parallel}\ket{\varphi_z}.
\end{equation}
As shown in Appendix~\ref{app:Q_parallel_SU2}, $Q_J^{\parallel}$ is
diagonal in the $J_z$ basis, its diagonal entries increase with $m$,
and therefore its largest eigenvalue occurs at $m=J$.  Hence the
optimal endpoint output is the guessed coherent state itself, and the
corresponding endpoint operation fidelity is
\begin{equation}
        F (I^{\parallel,j}_{\max})
        =
        (2J+1)
        \int_{-1}^{1}{dx\over2}
        \left({1+x\over2}\right)^{4J}
        =
        {2J+1\over4J+1}.
        \label{eq:Fparallel_maxI_arbitrary_j}
\end{equation}
Therefore
\begin{equation}
        D (I^{\parallel,j}_{\max })
        =
        1-F(I^{\parallel,j}_{\max })
        =
        {2J\over4J+1}
        =
        {4j\over8j+1}.
        \label{eq:Dparallel_maxI_arbitrary_j}
\end{equation}

\subsection{Endpoint comparison}

The parallel endpoint has closed formulas
(\ref{eq:Iparallel_max_arbitrary_j}) and
(\ref{eq:Dparallel_maxI_arbitrary_j}).  The antiparallel maximum
information is given by Eq.~(\ref{eq:Ianti_max_arbitrary_j}), while
the corresponding minimum disturbance is obtained from the finite
eigenvalue problem in Eq.~(\ref{eq:Danti_endpoint_block_formula}).

\begin{table}[t]
\caption{Endpoint information and disturbance for antiparallel and parallel
spin-coherent encodings.  The antiparallel disturbance is obtained by optimizing
the output state after the maximum-information POVM, i.e. from
Eq.~(\ref{eq:Danti_endpoint_block_formula}) after maximizing over
  all magnetic-number blocks.  For all displayed values the maximizing
  block is the $m=0$ block.}
\label{tab:endpoint_values_j}
\begin{ruledtabular}
\begin{tabular}{ccccc}
$j$ &
$I^{\rm anti}_{\max}$ &
  $D (I^{\rm anti}_{\max})$ &
$I^{\parallel}_{\max}$ &
 $D(I^{\parallel}_{\max})$ \\
\hline
$1/2$ & $0.788675$ & $0.315354$ & $0.750000$ & $0.400000$ \\
$1$   & $0.884773$ & $0.316957$ & $0.833333$ & $0.444444$ \\
$3/2$ & $0.923549$ & $0.311796$ & $0.875000$ & $0.461538$ \\
$2$   & $0.942636$ & $0.309422$ & $0.900000$ & $0.470588$ \\
$5/2$ & $0.953792$ & $0.308743$ & $0.916667$ & $0.476190$ \\
$3$   & $0.961168$ & $0.308782$ & $0.928571$ & $0.480000$ \\
$7/2$ & $0.966456$ & $0.309074$ & $0.937500$ & $0.482759$ \\
$4$   & $0.970455$ & $0.309428$ & $0.944444$ & $0.484848$ \\
$9/2$ & $0.973593$ & $0.309770$ & $0.950000$ & $0.486486$ \\
$5$   & $0.976125$ & $0.310081$ & $0.954545$ & $0.487805$
\end{tabular}
\end{ruledtabular}
\end{table}

\begin{figure}[t]
\begin{center}
\includegraphics[width=0.95\columnwidth]{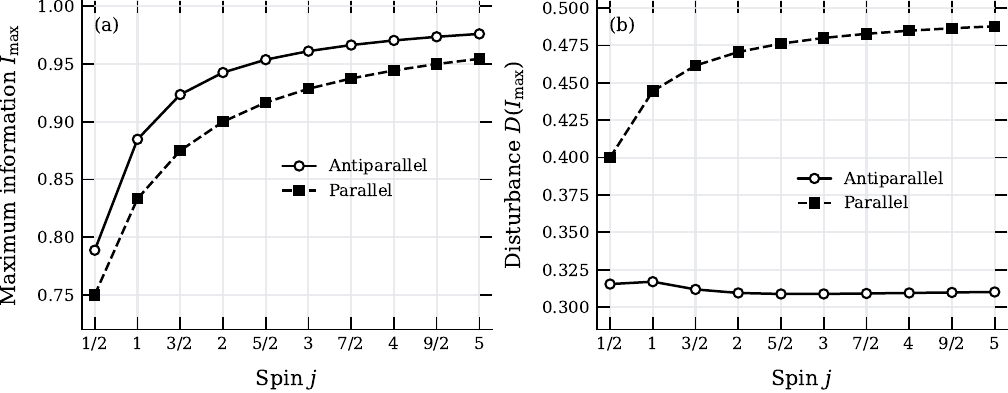}
\caption{Endpoint comparison for antiparallel and parallel
  spin-coherent encodings as a function of the spin $j$.  Solid curves
  with circles denote the antiparallel encoding, and dashed curves
  with squares denote the parallel encoding.  Panel (a) shows the
  maximum directional information.  Panel (b) shows the
  antiparallel endpoint disturbance obtained after optimizing the
  output state for the maximum-information POVM; it remains
  substantially below the parallel one.  The antiparallel curve is not
  strictly monotone; it displays a shallow minimum at intermediate
  spin.}
\label{fig:endpoint_encoding_comparison_PRA}
\end{center}
\end{figure}

Both endpoint information values approach unity as $j$ increases.  The endpoint
disturbances, however, behave differently.  For parallel spins,
\begin{equation}
        D(I^{\parallel,j}_{\max}) 
        =
        {4j\over8j+1}
        \longrightarrow {1\over2}.
        \label{eq:parallel_asymptotic_disturbance}
\end{equation}
For antiparallel spins the finite-matrix calculation gives a
disturbance close to $0.31$ over the range shown in
Table~\ref{tab:endpoint_values_j}, well below the corresponding
parallel value.  Thus, at the fully informative endpoint, the
antiparallel encoding gives both larger directional information and
smaller disturbance. The corresponding values of
Table~\ref{tab:endpoint_values_j} are also reported in
Fig. \ref{fig:endpoint_encoding_comparison_PRA}. The antiparallel
endpoint disturbance is not strictly monotone in $j$.  It has a
shallow minimum around $j=5/2$ in the range displayed, and then slowly
increases, remaining close to $0.31$ and well below the parallel
endpoint disturbance.  Thus the relevant robust conclusion is not
monotonicity, but the persistent separation between the antiparallel
and parallel endpoint disturbances.

The large-\(j\) limit of the antiparallel endpoint disturbance can also
be obtained analytically. As shown in Appendix \ref{asym} one has
\begin{eqnarray}
\lim_{j\to\infty}D(I_{\max}^{\rm anti,j})
        =
        8\sqrt2-11
        \simeq0.313708 .
\label{asy}
\end{eqnarray}
The shallow minimum observed at finite \(j\) is therefore not the
asymptotic value, but a finite-spin effect.

\paragraph*{Remark on the full tradeoff for higher spin.}
The endpoint formulas are simpler than the full tradeoff because the
maximum-information POVM is fixed and only the output state remains to be
optimized; the disturbance then follows from the finite eigenvalue problem in
Eq.~(\ref{eq:Danti_endpoint_block_formula}).  Away from the endpoint one must
optimize the full covariant Choi seed.  For antiparallel spin $j$ the dual
problem in Eq.~(\ref{eq:higher_dual}) contains $2j+1$ sector multipliers,
whereas the parallel spin-coherent benchmark is irreducible and has only one
trace constraint.

The rank structure of the optimal Choi seed is clarified in
Appendices~\ref{app:rank_one_simple_kernel} and~\ref{app:rank_bound}.
If the active kernel of the dual slack is one-dimensional,
Appendix~\ref{app:rank_one_simple_kernel} shows that its sector
weights are automatically correct and the optimal seed can be chosen
rank one.  Independently, Appendix~\ref{app:rank_bound} gives the
general low-rank guarantee
$\operatorname{rank}R_0^\star\le\lfloor\sqrt{2j+1}\rfloor$.  Thus rank
one is forced by convex geometry for $j=1/2$ and $j=1$, while for
higher spin it is a spectral property of the active dual kernel.  If
that kernel is genuinely degenerate, the optimal face is described by
Eq.~(\ref{eq:degenerate_kernel_seed}) with a positive matrix $G$
satisfying all sector normalizations.

It is useful to distinguish the coherent-product comparison studied here from
the ultimate direction-encoding problem with two spin-$j$ systems.  For
$j>1/2$, the state that maximizes the information over all pure two-spin states
is generally entangled and is not the antiparallel spin-coherent product state.
Equivalently, its amplitudes in the sectors
$\ell=0,\ldots,2j$ are not, in general, the Clebsch-Gordan coefficients in
Eq.~(\ref{eq:higher_c_l}).  For example, for two spin-one particles the
information-optimal fiducial state is
\begin{equation}
        \ket{\psi_z^{\rm opt}}
        ={\sqrt{10}\over6}\ket{0,0}
        +{1\over\sqrt2}\ket{1,0}
        +{\sqrt2\over3}\ket{2,0},
        \label{eq:spin_one_ultimate_opt_state}
\end{equation}
which gives
\begin{equation}
        I_{\rm opt}^{(j=1)}
        ={1\over2}+{\sqrt{15}\over10}
        \simeq 0.887298 .
        \label{eq:spin_one_ultimate_opt_info}
\end{equation}
The antiparallel spin-one coherent product state instead gives
\begin{equation}
        I_{\max}^{\rm anti,1}
        ={1\over2}+{\sqrt5\over15}+{\sqrt2\over6}
        \simeq0.884773 .
\end{equation}
The construction of information-optimal direction encodings can be
obtained by following the general procedure discussed in
Ref.~\cite{chir1}.  In the present work, however, we deliberately
compare two physically transparent separable encodings---parallel and
antiparallel spin-coherent pairs---and optimize the measurement back
action for those encodings.  To our knowledge, the
information-disturbance tradeoff for antiparallel spin-$j$ coherent
pairs, together with the corresponding endpoint comparison with
parallel spin-$j$ coherent pairs, has not been worked out previously.

\section{Conclusions}
\label{sec:conclusions}
We have formulated the optimal information-disturbance tradeoff for
the estimation of direction encoded in two spins as a covariant
Jamio\l{}kowski optimization problem.  The construction separates the
operational ingredients from the representation theory.  The
operational ingredients are the information functional, defined by the
average directional score, and the operation fidelity, defined by the
average overlap between the input state and the conditional output
state.  The representation theory determines the trace constraints on
the seed Choi operator: one independent constraint is obtained for
each irreducible sector appearing in the input orbit.

For two antiparallel spin-one-half particles the relevant representation
is $0\oplus1$.  The optimal supporting-line problem therefore has two dual
multipliers.  We wrote the dual slack operator explicitly and showed that the optimal
seed is supported on its zero eigenspace.  For generic interior points of the
spin-one-half curve this eigenspace is one-dimensional, giving a rank-one
filter generated from the kernel vector; if the zero eigenspace is degenerate,
the remaining task is the reduced positive feasibility problem within that
kernel.  The allowed filter has
scalar-vector mixing, which is precisely the structure required to exploit
the coherence responsible for the antiparallel advantage.  This also shows
why the optimal tradeoff is not generally obtained by simply interpolating
between doing nothing and performing a measurement followed by
repreparation.

At the maximum-information endpoint the POVM is the optimal covariant
orientation measurement.  The output state, however, remains a
variational parameter.  Optimizing it gives the antiparallel
spin-one-half disturbance, which is below both the disturbance of the
corresponding measure-and-reprepare antiparallel strategy and the
parallel-spin endpoint value.  Thus, antiparallel spins are better in
both senses: they provide more directional information and suffer less
minimum disturbance.

The arbitrary-spin endpoint analysis shows how this advantage is organized
for general spin-coherent signals.  For antiparallel spin-$j$ states the
orbit decomposes as $j\otimes j=\bigoplus_{\ell=0}^{2j}\ell$.  The
maximum information contains only nearest-neighbor coherences between these
sectors, while the corresponding minimum disturbance is the largest
eigenvalue of an explicit finite block-diagonal operator.  For the values
reported, the antiparallel endpoint information exceeds the parallel value
and the antiparallel endpoint disturbance remains substantially smaller
than the parallel disturbance.

Away from the maximum-information endpoint, the full higher-spin problem
is a genuine multiconstraint semidefinite program.  The dual slack contains
$2j+1$ independent multipliers, and its kernel may have dimension larger
than one; a feasible optimal Choi operator can then require a positive
combination of several kernel projectors satisfying all sector
normalizations.  This kernel-feasibility problem is absent in irreducible
parallel-spin benchmarks and is the main mathematical complication in the
complete higher-spin tradeoff.  The present formulation provides the
natural starting point for such calculations and for analytic certificates
of optimality in higher-dimensional cases.

\appendix

\section{Relation to a marginal-disturbance tradeoff for antiparallel spin-$1/2$ states}
\label{app:zhang_relation}
Reference~\cite{ZhangAntiTradeoff} studies a related information-disturbance problem for antiparallel spin-$1/2$ direction transmission, motivated by eavesdropping on reference-frame communication. The main difference is the disturbance functional. In that work the disturbance tests the survival of one output spin, through a fidelity of the form
\begin{equation}
        F_{\rm marg}=
        \int dg\sum_{r\mu}
        \Tr\!\left[
        A_{r\mu}\widetilde \rho _g A_{r\mu}^\dagger
        \left(|\psi _g \rangle\langle\psi _g |\otimes I\right)
        \right] .
        \label{eq:zhang_marginal_appendix}
\end{equation}
This is a marginal figure of merit: a measurement on the second spin
can leave the tested first-spin state undisturbed. Consequently the
zero-disturbance endpoint of that problem occurs at information $2/3$.

The present paper uses instead the full-state operation fidelity of
the complete two-spin signal,
\begin{equation}
        F_{\rm full}=
        \int d\vec n\sum_\mu
        \left|\langle\psi_z|A_{\vec n\mu}|\psi_z\rangle\right|^2 .
        \label{eq:full_fidelity_appendix}
\end{equation}
Here any nontrivial measurement on either spin generally disturbs the
complete input state, and the zero-disturbance endpoint is the
identity operation, with $I=1/2$. Thus the two analyses are related
but operationally distinct: Ref.~\cite{ZhangAntiTradeoff} studies a
marginal-disturbance tradeoff, whereas the present work studies the
full-state operation-fidelity tradeoff of the complete two-spin
signal. The comparison is mentioned here only to clarify the choice of
disturbance functional; the optimization carried out in the main text
concerns the full-state operation fidelity in
Eq.~(\ref{eq:full_fidelity_appendix}) and its natural extension to
arbitrary spin $j$.

\section{Endpoint output optimization at fixed covariant POVM}
\label{app:fixed_povm_Q}

This appendix gives the fixed-POVM construction used in the endpoint
calculations of the main text and writes it explicitly in terms of SU(2)
representation matrix elements.  The setting is the fully informative endpoint:
the covariant POVM has already been chosen to maximize the directional
information, and the only remaining variational freedom is the conditional
output state.

\subsection{General fixed-POVM construction}

Let the fixed endpoint POVM be rank one and covariant, with seed
$\ket{\eta_z}\bra{\eta_z}$.  For the outcome $\vec n=g\vec z$ we write
\begin{equation}
        \ket{\eta_{\vec n}}=V_g\ket{\eta_z} .
\end{equation}
At the endpoint, a covariant choice of pure output states is similarly
generated from a single fiducial vector $\ket{\varphi_z}$,
\begin{equation}
        \ket{\varphi_{\vec n}}=V_g\ket{\varphi_z} .
\end{equation}
For a true input $\ket{\psi_z}$, define
\begin{equation}
        \rho_z=\ket{\psi_z}\bra{\psi_z},
        \qquad
        p(g|z)=p(\vec n|z)
        =|\langle\eta_{\vec n}|\psi_z\rangle|^2 .
        \label{eq:app_likelihood_general}
\end{equation}
The endpoint operation fidelity at fixed POVM is
\begin{equation}
        F(\varphi)
        =\int d\vec n\,
        p(\vec n|z)
        |\langle\psi_z|V_g|\varphi_z\rangle|^2 =\bra{\varphi_z}Q\ket{\varphi_z},
        \label{eq:app_Q_quadratic_form}
\end{equation}
where
\begin{equation}
        Q=
        \int d\vec n\,
        p(\vec n|z)
        V_g^\dagger\rho_z V_g .
        \label{eq:app_Q_definition_compact}
\end{equation}
Thus $Q$ is the likelihood-weighted average of the true input projector,
rotated back to the fiducial output frame.  It does not determine the
maximum-information POVM; that POVM is already fixed and gives
information $I_{\max}$.  Instead, $Q$ solves the
remaining output-state optimization.  Since $Q\ge0$,
\begin{equation}
  F (I_{\max})
  =\lambda_{\max}(Q),
        \qquad
        D(I_{\max})=1-\lambda_{\max}(Q),
        \label{eq:app_Q_lambda_max}
\end{equation}
and the optimal fiducial output state is any normalized eigenvector associated
with $\lambda_{\max}(Q)$.

\subsection{Antiparallel spin-$j$ representation}
\label{app:Q_antiparallel_SU2}
We now derive the block formula used in Eq.~(\ref{eq:Danti_endpoint_block_formula}). We use the coupled basis $\ket{\ell,m}$ and the Wigner-matrix convention \cite{Edmonds}
\begin{equation}
        U_g^{(\ell)}\ket{\ell,m'}
        =\sum_m D^{(\ell)}_{m m'}(g)\ket{\ell,m} .
\end{equation}
The endpoint likelihood is the function $T_j(n_z)^2$ defined in Eqs.~(\ref{eq:endpoint_antiparallel_likelihood}) and (\ref{eq:endpoint_Tj}).  Its dependence only on $n_z=\cos\theta$ follows from the identity
\begin{equation}
        D^{(\ell)}_{00}(g)=d^{\ell}_{00}(\theta)=P_\ell(\cos\theta),
        \label{eq:app_D00_legendre}
\end{equation}
for the $m=0$ matrix element \cite{Edmonds}.

Substituting the decomposition of Eq.~(\ref{eq:higher_c_l}) into Eq.~(\ref{eq:Qj_endpoint_definition}) gives
\begin{equation}
\begin{split}
        \langle \ell,m|Q_j|\ell',m'\rangle
        &=c_\ell c_{\ell'}
        \int d\vec n\,
        T_j(n_z)^2
        D^{(\ell)}_{m0}(g^{-1})
        D^{(\ell')}_{0m'}(g) .
\end{split}
        \label{eq:app_Qj_D_matrix_elements}
\end{equation}
Writing $g$ in Euler angles, the azimuthal integral enforces $m=m'$.  Therefore $Q_j$ is block diagonal in the magnetic quantum number,
\begin{equation}
        Q_j=\bigoplus_m Q_j^{(m)} .
\end{equation}
The block with fixed $m$ has entries
\begin{equation}
        \left[Q_j^{(m)}\right]_{\ell\ell'}
        =c_\ell c_{\ell'}
        \int_{-1}^{1}{dx\over2}\,
        T_j(x)^2
        d^\ell_{m0}(\theta)d^{\ell'}_{m0}(\theta),
        \qquad x=\cos\theta,
        \label{eq:app_Qj_block_matrix}
\end{equation}
where $\ell,\ell'=|m|,|m|+1,\ldots,2j$.  Equation~(\ref{eq:Danti_endpoint_block_formula}) follows by maximizing the largest eigenvalue over these finite blocks.

\subsection{Parallel spin-$j$ representation}
\label{app:Q_parallel_SU2}

For the parallel endpoint we use the notation of Sec.~\ref{sec:endpoints_arbitrary_j}: the two-spin signal is equivalent to the spin-$J$ coherent state with $J=2j$ and fiducial vector $\ket{J,J}$.  Starting from Eq.~(\ref{eq:Qparallel_arbitrary_J}), the $J_z$-basis matrix elements are
\begin{equation}
        \langle J,m|Q_J^{\parallel}|J,m'\rangle
        =\int d\vec n\,p_J(n_z)
        D^{(J)}_{mJ}(g^{-1})D^{(J)}_{Jm'}(g) .
        \label{eq:app_Qparallel_D_matrix_elements}
\end{equation}
Since the likelihood $p_J(n_z)$ in Eq.~(\ref{eq:endpoint_parallel_likelihood}) depends only on the polar angle, the azimuthal integration gives
\begin{equation}
        \langle J,m|Q_J^{\parallel}|J,m'\rangle
        =\delta_{mm'}q_m^{\parallel} .
\end{equation}
Using the standard expression for reduced Wigner matrices \cite{Edmonds}
\begin{equation}
        \left|d^J_{mJ}(\theta)\right|^2
        ={2J\choose J+m}
        \left({1+\cos\theta\over2}\right)^{J+m}
        \left({1-\cos\theta\over2}\right)^{J-m},
\end{equation}
one obtains
\begin{equation}
        q_m^{\parallel}
        =(2J+1){2J\choose J+m}
        \int_{-1}^{1}{dx\over2}
        \left({1+x\over2}\right)^{3J+m}
        \left({1-x\over2}\right)^{J-m} .
        \label{eq:app_qm_parallel_integral}
\end{equation}
The beta integral gives
\begin{equation}
        q_m^{\parallel}
        =(2J+1){2J\choose J+m}
        {(3J+m)!(J-m)!\over(4J+1)!} .
        \label{eq:app_qm_parallel_closed}
\end{equation}
Furthermore,
\begin{equation}
        {q_{m+1}^{\parallel}\over q_m^{\parallel}}
        ={3J+m+1\over J+m+1}\ge1,
\end{equation}
so the largest eigenvalue is attained at $m=J$.  Hence
\begin{equation}
        F(I_{\max}^{\parallel,j})=q_J^{\parallel}
        ={2J+1\over4J+1},
        \qquad
        D(I_{\max}^{\parallel,j})={2J\over4J+1},
        \label{eq:app_parallel_endpoint_final}
\end{equation}
in agreement with Eqs.~(\ref{eq:Fparallel_maxI_arbitrary_j}) and (\ref{eq:Dparallel_maxI_arbitrary_j}).  For two parallel spin-$1/2$ particles, $J=1$ and
\begin{equation}
        Q_{J=1}^{\parallel}
        =\operatorname{diag}\left({1\over10},{3\over10},{3\over5}\right)
\end{equation}
in the ordered basis $\{\ket{1,-1},\ket{1,0},\ket{1,1}\}$.

\section{Rank-one optimality from a one-dimensional dual kernel}
\label{app:rank_one_simple_kernel}

We first record a sufficient condition for rank-one optimality of the covariant
Choi seed.  The general low-rank guarantee is proved in
Appendix~\ref{app:rank_bound}; the present result identifies the common case in
which the optimal seed is actually rank one.

Use the notation of Sec.~\ref{sec:higher_spin}: the sector constraints are
\begin{equation}
        \Tr(A_\ell R_0)=d_\ell,
        \qquad d_\ell=2\ell+1,
        \qquad \ell=0,\ldots,2j,
        \label{eq:simple_kernel_constraints}
\end{equation}
and the supporting-line objective is $C_\mu=R_F^{(j)}+\mu R_I^{(j)}$.  The
dual slack is
\begin{equation}
        K_\mu(\lambda)=\sum_{\ell=0}^{2j}\lambda_\ell A_\ell-C_\mu .
        \label{eq:simple_kernel_slack}
\end{equation}
Let $\lambda^\star$ be an optimal dual point and suppose that the lowest
eigenvalue of $K_\mu(\lambda^\star)$ is simple.  The dual constraint is active
at optimum, so this eigenvalue is zero.  Let $\dvec{X_\mu}$ be the normalized
kernel vector 
\begin{equation}
        K_\mu(\lambda^\star)\dvec{X_\mu}=0,
        \qquad
        \dbra{X_\mu} X_\mu \rangle\!\rangle =1 .
\end{equation}
Since the active lowest eigenvalue is assumed simple, the matrix
constraint \(K_\mu(\lambda)\ge0\) is locally equivalent to the scalar
constraint \(k_{\min}(\lambda)= \lambda_{\min}[K_\mu(\lambda)]\ge0\).  The KKT stationarity condition
for minimizing \(f(\lambda)=\sum_\ell d_\ell\lambda_\ell\) under this
active constraint gives $ d_\ell = t\,\frac{\partial
  k_{\min}}{\partial\lambda_\ell}$ for some multiplier \(t>0\).
First-order perturbation theory for a
simple eigenvalue yields \cite{Kato}
\begin{eqnarray}
\frac{\partial k_{\min}}{\partial\lambda_\ell}
        =
        \dbra{X_\mu}
        {\partial K_\mu\over\partial\lambda_\ell}
        \dvec{X_\mu}
=
        \langle\!\langle X_\mu|A_\ell|X_\mu\rangle\!\rangle ,
\end{eqnarray}
and therefore
\begin{eqnarray}
        d_\ell
        =
        t\,\langle\!\langle X_\mu|A_\ell|X_\mu\rangle\!\rangle .
\end{eqnarray}
Summing over \(\ell\) and using \(\sum_\ell A_\ell=I\) and
\(\langle\!\langle X_\mu|X_\mu\rangle\!\rangle=1\) gives
\(t=\sum_\ell d_\ell=(2j+1)^2\). Hence
\begin{equation}
        \dbra{X_\mu}A_\ell\dvec{X_\mu}
        ={2\ell+1\over(2j+1)^2},
        \qquad \ell=0,\ldots,2j .
        \label{eq:simple_kernel_weights}
\end{equation}
Therefore
\begin{equation}
        R_0^\star(\mu)=(2j+1)^2\dvec{X_\mu}\dbra{X_\mu}
        \label{eq:simple_kernel_rankone_seed}
\end{equation}
satisfies all sector constraints in Eq.~(\ref{eq:simple_kernel_constraints}).
Moreover $K_\mu(\lambda^\star)R_0^\star=0$, so complementary slackness holds;
Eq.~(\ref{eq:simple_kernel_rankone_seed}) is therefore optimal.

If the zero eigenspace is degenerate, the same reasoning gives a subgradient
condition rather than a single eigenvector condition.  The required sector
weight vector then lies in the convex hull generated by projectors onto the
active kernel.  This guarantees an optimal seed of the form
Eq.~(\ref{eq:degenerate_kernel_seed}), but not necessarily a single kernel
vector with the weights in Eq.~(\ref{eq:simple_kernel_weights}).  Rank-one
optimality in a degenerate active kernel is therefore a reduced
kernel-feasibility question.

Numerically, all higher-spin branches tested in this work up to $j=5$ have a
one-dimensional active kernel in the axially symmetric block.  Whenever this occurs,
Eq.~(\ref{eq:simple_kernel_weights}) follows automatically from stationarity,
so the sector-weight check is a consequence of the dual certificate rather than
an independent fit.

\section{Rank bound for optimal covariant Choi seeds}
\label{app:rank_bound}

We now prove a complementary rank-reduction bound that does not assume
a one-dimensional dual kernel.  It uses only positivity and the number
of independent sector-normalization constraints.

Consider the feasible set defined by Eq.~(\ref{eq:simple_kernel_constraints}).
It is compact because
\begin{equation}
        \sum_{\ell=0}^{2j}A_\ell=I_{\rm in}\otimes I_{\rm out},
        \qquad
        \Tr R_0=\sum_{\ell=0}^{2j}d_\ell=(2j+1)^2 .
        \label{eq:rank_bound_trace}
\end{equation}
Hence a linear supporting-line objective has an optimal extreme point.
Let $R_0$ be such an extreme point, let $r=\operatorname{rank}R_0$, and let
$P$ project onto its support.  The real vector space
${\cal H}_P=\{H=H^\dagger:\ H=PHP\}$ has dimension $r^2$.  The linear map
\begin{equation}
        {\cal L}:{\cal H}_P\longrightarrow\mathbb{R}^{2j+1},
        \qquad
        {\cal L}(H)_\ell=\Tr(A_\ell H)
        \label{eq:rank_bound_linear_map}
\end{equation}
has a nonzero kernel if $r^2>2j+1$.  A nonzero $H$ in this kernel obeys
\begin{equation}
        \Tr(A_\ell H)=0,
        \qquad \ell=0,\ldots,2j .
        \label{eq:rank_bound_null_direction}
\end{equation}
Since $H$ is supported on the support of $R_0$, the operators
$R_0\pm\epsilon H$ remain positive for sufficiently small $\epsilon>0$ and,
by Eq.~(\ref{eq:rank_bound_null_direction}), satisfy all sector constraints.
This would express $R_0$ as a nontrivial convex combination of two feasible
points, contradicting extremality.  Therefore every extreme feasible point
satisfies
\begin{equation}
        r^2\le 2j+1 .
        \label{eq:rank_bound_main}
\end{equation}
Consequently, for every $\mu$ there exists an optimal antiparallel Choi seed
with
\begin{equation}
        \operatorname{rank}R_0^\star
        \le
        \left\lfloor\sqrt{2j+1}\right\rfloor .
        \label{eq:rank_bound_final}
\end{equation}
Thus this bound forces rank one for $j=1/2$ and $j=1$, while for
$j\ge3/2$ it gives only low-rank existence.  The
one-dimensional-kernel criterion of
Appendix~\ref{app:rank_one_simple_kernel} explains when the optimal
seed can nevertheless be chosen rank one.

For the parallel benchmark the representation is irreducible.  There is only
one constraint, $\Tr R_0=2J+1$, so the feasible set is a scaled density-matrix
set and its extreme points are rank-one projectors.  Hence the parallel-spin
tradeoff always admits an optimal seed
\begin{equation}
        R_0^\parallel(\mu)
        =(2J+1)\dvec{\chi_\mu}\dbra{\chi_\mu},
\end{equation}
where $\dvec{\chi_\mu}$ is a largest-eigenvalue vector of
$R_F^\parallel+\mu R_I^\parallel$.

\section{Large-spin asymptotics of the antiparallel endpoint}
\label{asym}
This appendix derives the large-$j$ limit of the maximum-information
endpoint disturbance for antiparallel spin-coherent pairs.  The starting
point is the finite block formula in Eq.~(\ref{eq:Qj_m_block}).  We set
$N=2j$ and take the joint scaling limit
\begin{equation}
        \ell=\sqrt N\,s,\qquad
        \ell'=\sqrt N\,t,\qquad
        \theta={u\over\sqrt N},
        \label{eq:asym_scaling}
\end{equation}
with $s,t,u$ fixed as $N\to\infty$.  This is the relevant scaling
because the maximum-information likelihood becomes concentrated in an
angular region of width $N^{-1/2}$ around the true direction, while the
Clebsch-Gordan weights in Eq.~(\ref{eq:higher_c_l}) are concentrated on
angular momenta of order $\sqrt N$.

First, the factorial expression in Eq.~(\ref{eq:higher_c_l}) gives, by
Stirling expansion,
\begin{equation}
        c_\ell^2
        =(2\ell+1){(N!)^2\over (N-\ell)!(N+\ell+1)!}
        \sim {2s\over\sqrt N}\,e^{-s^2},
        \label{eq:asym_c_l}
\end{equation}
and therefore
\begin{equation}
        c_\ell c_{\ell'}
        \sim {2\sqrt{st}\over\sqrt N}
        \exp\!\left[-{s^2+t^2\over2}\right].
        \label{eq:asym_c_l_product}
\end{equation}
The Legendre polynomial and Wigner $d$-matrix elements have the
standard Bessel limits \cite{WatsonBessel,OlverNIST}
\begin{equation}
        P_\ell\!\left(\cos {u\over\sqrt N}\right)
        \simeq J_0(su),
        \qquad
        d^\ell_{m0}\!\left({u\over\sqrt N}\right)
        \simeq (-1)^m J_m(su),
        \label{eq:asym_bessel_limits}
\end{equation}
for fixed $m$.  Using Eq.~(\ref{eq:asym_c_l}), the endpoint amplitude
$T_j$ in Eq.~(\ref{eq:endpoint_Tj}) satisfies
\begin{equation}
        T_j\!\left(\cos {u\over\sqrt N}\right)
        \sim \sqrt N\int_0^\infty
        2s e^{-s^2/2}J_0(su)\,ds       
       =2\sqrt N\,e^{-u^2/2},
        \label{eq:asym_Tj}
\end{equation}
where the last equality follows from the Hankel transform of a
Gaussian. Moreover, since $x=\cos\theta$,
\begin{equation}
        {dx\over2}\simeq {u\,du\over 2N}.
        \label{eq:asym_measure}
\end{equation}
Substitution of Eqs.~(\ref{eq:asym_c_l_product})-(\ref{eq:asym_measure})
into Eq.~(\ref{eq:Qj_m_block}) gives
\begin{equation}
        \left[Q_j^{(m)}\right]_{\ell\ell'}
        \sim {1\over\sqrt N}K_m(s,t),
        \qquad s={\ell\over\sqrt N},\quad t={\ell'\over\sqrt N},
        \label{eq:asym_Q_to_kernel}
\end{equation}
where
\begin{equation}
\begin{split}
        K_m(s,t)
        &= {2\sqrt{st}}\,
        \exp\!\left[-{s^2+t^2\over2}\right]
        2\int_0^\infty u e^{-u^2}J_m(su)J_m(tu)\,du  \\
        &=2\sqrt{st}\,
        \exp\!\left[-{3\over4}(s^2+t^2)\right]
        I_m\!\left({st\over2}\right).
\end{split}
        \label{eq:asym_kernel_K}
\end{equation}
In the second line we used a standard integral for Bessel
functions \cite{WatsonBessel,GradshteynRyzhik}.  The finite-dimensional
eigenvalue equation for $Q_j^{(m)}$ therefore converges to the integral
equation
\begin{equation}
        \int_0^\infty K_m(s,t)f(t)\,dt=\lambda f(s).
        \label{eq:asym_integral_equation}
\end{equation}
It remains to diagonalize the kernel.  Put \(x=s^2\) and write
\(f(s)=x^{1/4}\phi(x)\).  Equation~(\ref{eq:asym_integral_equation}) is then
equivalent to
\begin{equation}
        \int_0^\infty L_m(x,y)\phi(y)\,dy=\lambda\phi(x),
        \label{eq:asym_L_equation}
\end{equation}
with
\begin{equation}
        L_m(x,y)=
        \exp\!\left[-{3\over4}(x+y)\right]
        I_m\!\left({\sqrt{xy}\over2}\right).
        \label{eq:asym_L_kernel}
\end{equation}
We diagonalize this kernel by using the generating formula
for Laguerre polynomials \cite{pleg,OlverNIST}.  In the form needed here,
for \(0<\zeta<1\) and \(\alpha>-1\),
\begin{equation}
\begin{split}
&\sum_{n=0}^{\infty}
        {n!\over\Gamma(n+\alpha+1)}
        L_n^{(\alpha)}(X)L_n^{(\alpha)}(Y)\zeta^n    \\
&\quad =
        {1\over 1-\zeta}
        \exp\!\left[-{\zeta(X+Y)\over1-\zeta}\right]
        (XY\zeta)^{-\alpha/2}
        I_\alpha\!\left({2\sqrt{XY\zeta}\over1-\zeta}\right).
\end{split}
        \label{eq:hille_hardy}
\end{equation}
We apply Eq.~(\ref{eq:hille_hardy}) with
\[
        \alpha=m,\qquad X=\beta x,\qquad Y=\beta y.
\]
Define the normalized Laguerre functions
\begin{equation}
        \Phi_n^{(m)}(x)=
        \left({\beta^{m+1}n!\over\Gamma(n+m+1)}\right)^{1/2}
        x^{m/2}e^{-\beta x/2}L_n^{(m)}(\beta x).
        \label{eq:asym_laguerre_functions}
\end{equation}
Then Eq.~(\ref{eq:hille_hardy}) gives
\begin{equation}
\sum_{n=0}^{\infty}
        \zeta^n
        \Phi_n^{(m)}(x)\Phi_n^{(m)}(y)       \quad =
        {\beta\over1-\zeta}
        \zeta^{-m/2}
        \exp\!\left[
        -{\beta(1+\zeta)\over2(1-\zeta)}(x+y)
        \right]
        I_m\!\left(
        {2\beta\sqrt{\zeta xy}\over1-\zeta}
        \right).
        \label{eq:hille_hardy_normalized}
\end{equation}
We now match Eq.~(\ref{eq:hille_hardy_normalized}) with the kernel
\(L_m(x,y)\) in Eq.~(\ref{eq:asym_L_kernel}).  This requires
\begin{equation}
        {\beta(1+\zeta)\over2(1-\zeta)}={3\over4},
        \qquad
        {2\beta\sqrt{\zeta}\over1-\zeta}={1\over2}.
        \label{eq:asym_matching}
\end{equation}
The solution with \(0<\zeta<1\) is
\begin{equation}
        \beta=\sqrt2,\qquad
        \zeta=17-12\sqrt2 .
        \label{eq:asym_zeta_solution}
\end{equation}
Therefore
\begin{equation}
        L_m(x,y)=
        {1-\zeta\over\beta}
        \zeta^{m/2}
        \sum_{n=0}^{\infty}
        \zeta^n
        \Phi_n^{(m)}(x)\Phi_n^{(m)}(y).
        \label{eq:asym_L_spectral}
\end{equation}
The eigenvalues of the integral operator are consequently
\begin{equation}
        \lambda_n^{(m)}
        =
        {1-\zeta\over\beta}\,
        \zeta^{m/2+n}.
        \label{eq:asym_laguerre_eigenvalues}
\end{equation}
The largest eigenvalue is obtained for \(m=0\) and \(n=0\):
\begin{equation}
        \lambda_{\max}=12-8\sqrt2 .
        \label{eq:asym_largest_eigenvalue}
\end{equation}
Consequently,
\begin{equation}
        \lim_{j\to\infty}F(I_{\max}^{\rm anti,j})
        =12-8\sqrt2,
        \qquad
        \lim_{j\to\infty}D(I_{\max}^{\rm anti,j})
        =8\sqrt2-11 .
        \label{eq:asym_final_result}
\end{equation}


\begin{thebibliography}{99}
\bibitem{nota} Although this issue does not arise in the present analysis, equivalent irreducible components cannot in general be ignored. They can be relevant for ultimate optimizations, for example in reference-frame transmission \cite{chiri2} or squeezing estimation \cite{sque}.
\end{thebibliography}
\end{document}